\documentclass[ASNA,twocolumn]{USG} %twocolumn
\usepackage{anyfontsize} %

\usepackage{colortbl}   % For coloring table lines
\usepackage{xcolor}     % For color definitions
\usepackage{adjustbox}
\usepackage{steinmetz}

\makeatletter
\let\USGtitlefont\titlefont
\renewcommand{\titlefont}{\USGtitlefont\centering}
\let\USGAuthorfont\Authorfont
\renewcommand{\Authorfont}{\USGAuthorfont\centering}
\renewcommand{\orchid}[1]{\href{#1}{\tikz[baseline=-0.55ex]{\node[circle,fill=green!55!black,inner sep=0pt,minimum size=0.85em]{\color{white}\sffamily\bfseries\tiny iD};}}}
\makeatother

\historydates{}

\begin{document}
\title{An Empirical Study on CVE-based Vulnerabilities Propagation from one System to another System (IntraProject vs InterProject)}
\author[1]{Junaid Akram}[https://orcid.org/0000-0003-1752-6124]
\author[2]{Ali Hassan}
\author[3]{Danish Vasan}[https://orcid.org/0000-0002-7693-1042]

\authormark{Junaid Akram \textsc{et al.}}
\titlemark{An Empirical Study on CVE-based Vulnerabilities Propagation from one System to another System (IntraProject vs InterProject)}

\address[1]{\orgdiv{School of Computer Science, University of Nottingham Ningbo, China (UNNC)}}

\address[2]{\orgdiv{Software Development and Operations, Punjab Sahulat Bazaars Authority, Lahore, Pakistan}}

\address[3]{\orgdiv{Research Center for Intelligent Secure Systems, King Fahd University of Petroleum \& Mineral, Dhahran, Saudi Arabia}}

\corres{Junaid Akram  (\email{dr.junaidakram@outlook.com}) ~}

\keywords{Software Evolution | Software Security | Software Maintenance | Software Reuse | Vulnerability Detection | Vulnerability Propagation | CVE}

\abstract[ABSTRACT]{Vulnerabilities in applications, including browsers, media players, online services and document readers, etc. are often exploited and cause a serious damage. The vulnerabilities could exist in your system maybe because of some coding errors, design oversights, outdated software, AI generated codes or maybe transferred from another system during development (e.g., code clones). The propagation of software vulnerabilities is a critical concern in cybersecurity. This study examines the root cause problem of vulnerabilities propagation with experiments. We analyzed the propagation of CVE (Common Vulnerabilities and Exposures) based vulnerabilities within and between different software projects (IntraProject vs. InterProject). We performed several experiments on CVE based vulnerability dataset (retrieved from NVD source) by using our vulnerability detection tool (SQVDT). In the results, we discovered similar patterns of vulnerable code in several very famous recently used software, including latest version of Linux, Android operating systems, and in thousands of open source projects available on GitHub. We found several very high severity level (CVSS) vulnerabilities that propagated from one system to another system, just because of copy/paste (reuse) the code. Furthermore, we performed an empirical evaluation and verification of these vulnerable patterns that propagated to other systems by also considering the frequency of their propagation. Finlay, we recommended the impact of these vulnerabilities propagation on the interconnected systems.}

\maketitle

\section{Introduction}
Recently, a lot of efforts and investments have been devoted to software security, especially toward minimizing the impact of software vulnerabilities. Source code security assurance is a process in which we make sure that the product design protects the resources and data inside it. As the number of threats targeting software systems continues to rise, ensuring their security becomes crucial. This security assurance is closely tied to the quality of the source code. In the domain of information technology, software security assurance plays an important role in maintaining business continuity, safeguarding personal data, enhancing national security, fortifying cybersecurity, internet security, and homeland security. Therefore, identifying and patching the software vulnerability before it is used for exploit becomes even more important. By addressing security concerns at the code level, developers can enhance the resilience of software systems. This proactive approach helps safeguard against potential threats and ensures a more robust and secure digital landscape.

Vulnerability exploits serve as a significant vectors for malware infection. For instance, a CVE-2017-5638 bug was disclosed in March 2017, but hackers continued to exploit it until May 2017 against Equifax \cite{lawler2017equifax}. During this breach, hackers stole personal information from over 143 million users in the US, including their names, birth details, and social security numbers \cite{avery2018formally}. Another critical bug, CVE-2018-1274, emerged on April 10, 2018 \cite{nvdCVE20181274}, it caused a denial-of-service (DoS) attack on authenticated remote users by straining memory and CPU resources. Fortunately, it was fixed on April 17, 2018. Another very critical (CVSS score: 8.0) security vulnerability CVE-2024-4835 \footnote{https://nvd.nist.gov/vuln/detail/CVE-2024-4835} has recently published, it is a cross-site scripting (XSS) flaw, which could have allowed attackers to inject malicious code into the GitLab web pages. GitLab has already released patches (fixes) to address this vulnerability in May 2024. This vulnerability allows a hacker to \textit{Remote Code Execution (RCE)}, which means an attacker could potentially take complete control of your device if it’s exploited. Despite these efforts, many users are still remain unaware of such vulnerabilities. Some non-technical users simply ignore the security updates (patch files). Additionally, developers sometimes reuse code fragments from other software systems without thoroughly checking and scanning their code against these vulnerabilities.

Some vulnerabilities affect you, if you are using third-party opens source components. As modern software applications, including commercial ones, heavily rely on Open-Source Software (OSS) components. These OSS components constitute approximately 90\% of the software products in the market \cite{syvertsen2011selection}. However, this widespread use of open source introduces a significant security challenge to the software community. Developers often depend on outdated versions of libraries that may contain several severe types of software vulnerabilities. For example, a vulnerability of CVE-id (CVE-2014-0160), which is a heart-bleed ``OpenSSL" vulnerabilty, that has affected several systems including, websites, web-applications, servers, and the operating systems. The affected systems were using this library as whole or a part of it \cite{kim2017vuddy}. Vulnerability detection and the creation of patch (fixes) files have significantly improved our ability to swiftly address security threats in the software systems, as an automatic exploitation becomes feasible through these patch files \cite{brumley2008automatic, Akram2019IBFET, Akram2019how}. Patches applied to vulnerable code play a crucial role in ensuring security measures \cite{alhazmi2007measuring, gorbenko2017experience, mumtaz2019rsa, guo2016research}. Interestingly, research indicates that whenever a patch is released, an attacker (often referred to as a hacker) can quickly leverage it to pinpoint the exact location of the buggy code and automatically create an exploit \cite{Akram2019VCIPR, akram2021sqvdt}. In recent years, open-source software has contributed to an increased rate of vulnerabilities, as many software developers copy and paste code without evaluating or testing the code \cite{akram2021sqvdt}.

\begin{figure*}[htbp]
  \centering
  \includegraphics[scale=.23]{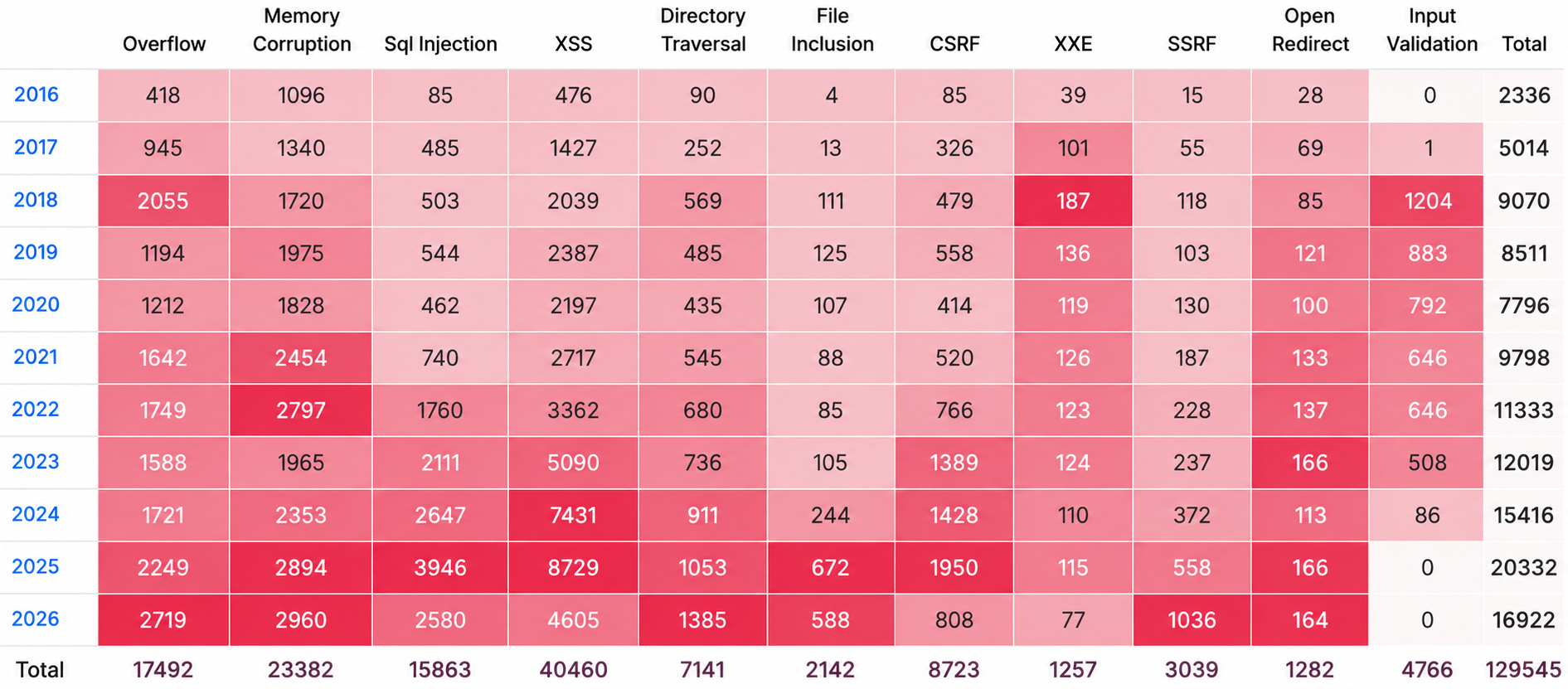}
  \caption{Vulnerabilities by Types / Year (data source: CVEdetails.com), from 2016 to 2026, access date: August 2026 \label{fig:VulYear}}
\end{figure*}

Code clones are also playing a crucial role in vulnerabilities propagation, and these clones always bring a lot of troubles to maintain the software security \cite{ng2014android}. Previous research shows that about 7\% to 23\% of source code is cloned in large systems \cite{juergens2009code}, and this cloned code could be a vulnerable or buggy code. Sheneamer et al. \cite{sheneamer2016survey} claims that almost 22\% of Linux code is cloned code, which could be cloned from previous versions of Linux or from other open source software. Some very famous Android vendors are using Linux-kernel in their operating systems \cite{akram2021sqvdt}, and there can be a possibility that this cloned kernel is from an old version of Linux, which could be vulnerable and unpatched. Although Linux has repaired this vulnerability through a patch file in the original code but remained unpatched in the old versions. Therefore, we can say that code cloning is also a major cause of vulnerability propagation from one system to another system.

Another very important and modern factor that could involves vulnerability propagation is the use of generative-AI tools to generate code and use it in our systems \cite{kapitsaki2024generative}. GenAI started to become a cybersecurity concern in 2023 and it is going to become a much bigger issue in 2024 and beyond \cite{techtarget}. The advent of generative-AI tools has started in a new era of software development, which has accelerated the development cycles and productivity. However, this advancement has also introduced an unforeseen factor of vulnerability propagation. As the AI models are trained on the bulk of code (mostly open source), the generated code could have some security loopholes, such as \textit{Buffer Overflows, SQL Injection, or Cross-Site Scripting} in the code that can become a threat to the system \cite{cotroneo2024vulnerabilities}. These vulnerabilities, can be exploited and the system could be compromised, which low down the overall system reliability. To overcome these risks, the developers and software engineers must be equipped with the latest knowledge and tools to assess or evaluate (manually/automatically) the generated code fragments thoroughly before they use in their own systems. Code reviews, code testing, and vulnerability scanning should become an integrated parts of the development process.

\textbf{Motivation:} The main motivation to conduct this research is shown in \figurename~\ref{fig:VulYear}. The figure illustrates the prevalence and evolution of various types of software vulnerabilities over the years, categorized by common attack vectors such as \textit{Overflow, Memory Corruption, SQL Injection, XSS (Cross-Site Scripting)}, and others. The data highlights the persistent and evolving nature of security threats in software, emphasizing the critical need for ongoing research and innovation in vulnerability detection and mitigation. The rising trends and total counts serve as a clear motivation for conducting this research to address these persistent security challenges, especially as new vulnerabilities continue to emerge. Additionally, the following concerns motivate us to find the root cause reasons behind the vulnerability propagation from one system to another system.
\begin{itemize}
  \item The increase in software re-usability by simply cloning the code, which led to a big threat.
  \item A lot of old vulnerabilities found in some latest open source software \cite{akram2021sqvdt}.
  \item Difficulties to perform an audit on open source software before use their code to quick the process of development.
  \item Software dependency to use some external components including libraries, utilities, and programs, which can be faulty or vulnerable. An analysis shows that there are vulnerabilities in almost 0.62\% Maven projects \cite{alqahtani2016tracing}.

  \item The annual average cost of cybercrime is predicted to hit more than 23 trillion dollar in 2027 \cite{techtarget}
\end{itemize}

As modern software systems are under a major security threats due to the increasing trend of using code clones for software re-usability. This often results in migration of old vulnerabilities into new releases of such open source software. For example, a vulnerability CVE-2013-0310 \footnote{https://nvd.nist.gov/vuln/detail/CVE-2013-0310} is a \textit{denial of services} was first discovered in Linux kernel before version 3.4.8, in year 2013. In year 2020, the same vulnerability was discovered in several open source software available on GitHub \cite{akram2021sqvdt}, which shows that people were still using this vulnerable code. In addition, promptness to speed up development may make it difficult to examine sufficiently an open-source program prior to merging it to the code. Furthermore, recent AI-based code generation tools such as ChatGPT have presented another challenge: introducing potentially unsafe code into our systems. This paper studies how vulnerabilities spread within and across the software projects aiming at developing an effective mitigation strategies and improving general software security. Based on our analysis of thousands of vulnerabilities that were detected by our tool (SQVDT), including vulnerabilities in almost 152K open source projects, latest version of Linux, and Android operating systems from different vendors; we presents our findings on the root causes of vulnerability propagation from one system to another system. Following are the main contributions in this paper.

\begin{itemize}
  \item We perform an empirical analysis of thousand of vulnerability patches published against different CVEs.
  \item Develop a crawler that extracted thousands of CVE based vulnerable source code fragments by tracing their patches.
  \item We detect thousands of vulnerable patterns that propagated from one system to another system by using our tool SQVDT.
  \item We identified and categorized the propagation mechanisms of these CVE-based vulnerabilities.
  \item Compare the frequency and impact of IntraProject, InterProject, Library-used.
  \item Recommendations to stop the vulnerabilities propagation to other systems.
\end{itemize}

\section{Background and Preliminaries}

\subsection{Terminologies}
Here are some terminologies, that are used in this paper.\\
\textbf{Vulnerability: }It is a weakness or a flaw in software system, which allows an attacker (hacker) to overcome a system's security.  \\
\textbf{NVD:} NVD (National Vulnerability Database) \footnote{https://nvd.nist.gov/} is the United States (government) standards based vulnerability repository. It includes security references, flaws, security misconfigurations, product names and its metrics.\\
\textbf{CVE: }It is the Common Vulnerabilities Exposures \footnote{https://cve.mitre.org/} system, which provides a reference CVE-IDs for publicly known vulnerabilities, including a unified identification number, description, and a reference number for known vulnerabilities.  \\
\textbf{CWE: }Weakness Enumeration is a category system for hardware and software weaknesses and vulnerabilities.\\
\textbf{CVSS:} It provides a principal characteristics of a vulnerability and produce a numerical score reflecting its severity (level 0 - 10).\\
\textbf{Code Clones: }A copied code fragment from other systems is called a code clone. There are four different types of code clones \cite{akram2018droidcc}.\\
\textbf{SQVDT:} Scalable Quantitative Vulnerability Detection
Technique.\\
\textbf{Generative-AI: }Generative artificial intelligence is capable of generating text, images, videos, or other data using generative models, often in response to prompts.\\
\textbf{Patch File:} It is a series of instructions in the source code that describes how to change a file or a set of files in an application.\\
\textbf{Vulnerability Propagation:} The process by which a security vulnerability in one piece of software spreads or affects other systems, either within the same project or across different projects.\\
\textbf{IntraProject Propagation:} The spread of a vulnerability within the same software project, typically through code reuse, module sharing, or shared libraries. \\
\textbf{InterProject Propagation:} The transmission of vulnerabilities from one software project to another, often through the reuse of open-source code, third-party libraries, or cloned codebases.\\
\textbf{Software Reusability:} The practice of using existing code in new projects, which can lead to the unintentional propagation of vulnerabilities if the reused code contain flaws.

\begin{figure*}
  \centering
  \includegraphics[scale=.23]{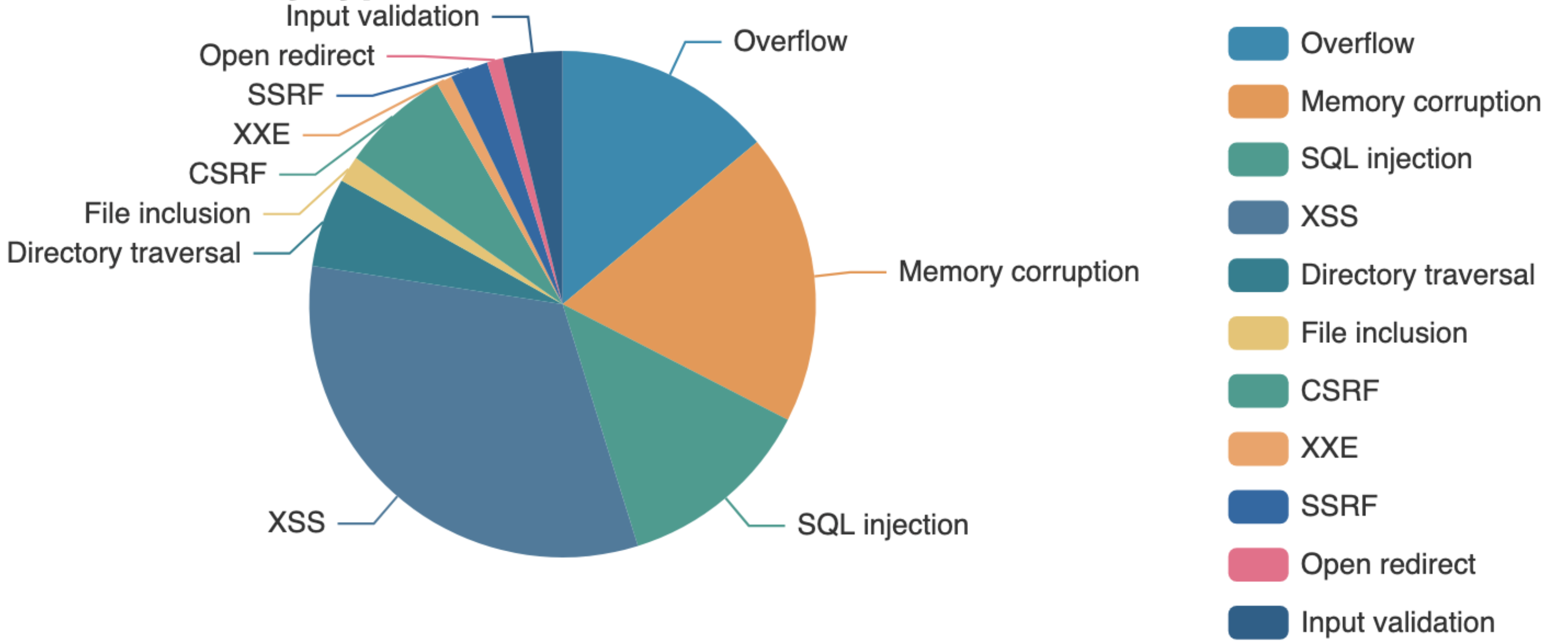}
  \caption{Vulnerabilities by types (data source: CVEdetails.com), from 2016 to 2026, access date: August 2026 \label{fig:VulTypes}}
\end{figure*}

\subsection{Software Vulnerabilities and CVEs}
Software vulnerabilities are the weaknesses or flaws in a system's design, implementation, or configuration that can be exploited by attackers to compromise the system's security. There are different types of vulnerabilities in the source code, i.e. \textit{injection vulnerabilities, buffer overflow, sensitive data exposure, broken authentication, XSS (Cross-Site Scripting)}, etc. \figurename ~\ref{fig:VulTypes} visually represents the distribution of various types of software vulnerabilities from 2016 to 2026. It illustrates that \textit{overflow}, \textit{memory corruption}, and \textit{XSS } being the most prevalent. \textit{Overflow} vulnerabilities involve writing more data to a memory location than it can hold, are the most common due to frequent programming errors in managing memory boundaries. \textit{Memory corruption}, the second most common type, occurs when unexpected changes to memory content lead to erratic behavior, often exploited in attacks like buffer overflows. XSS is another most occurred vulnerability, in which malicious scripts are injected into benign and trusted websites. XSS attacks occur when an attacker uses a web application to send malicious code, generally in the form of a browser side script, to a different end user \footnote{https://owasp.org/www-community/attacks/xss/}. These vulnerabilities can lead to unauthorized access, data breaches, system disruptions, and other security incidents. To systematically identify, categorize, and track these vulnerabilities, the Common Vulnerabilities and Exposures (CVE) system was established. The CVE system provides unique identifiers (CVE IDs) for publicly known vulnerabilities, serving as a standardized reference across the cybersecurity community, managed by the MITRE \footnote{https://cve.mitre.org/} Corporation.

\subsection{Importance of Vulnerable Code and Patches}
The analysis of vulnerable source code plays a crucial role in understanding the root causes of security flaws and developing effective mitigation strategies. By examining a specific code that led to a vulnerability, researchers and developers can identify patterns, common mistakes, and the underlying factors that contribute to insecure a software. This analysis not only aids in fixing current vulnerabilities but also in preventing similar issues in future software development. However, obtaining a vulnerable source code is often challenging. While the CVE database provides detailed descriptions of vulnerabilities, it does not always include direct links to the corresponding source code or patch files. This limitation can hinder efforts to study and address vulnerabilities comprehensively. Additionally, many vulnerabilities are found in proprietary or closed-source software, where access to the source code is restricted, further complicating vulnerability analysis.

\subsection{Vulnerability Analysing and Crawling}
To overcome the challenge of acquiring vulnerable source code, we resort to automated methods, such as web crawlers. Web crawlers are automated scripts or programs designed to systematically browse the internet and retrieve specific types of information (vulnerable code in our case). By targeting specific reference sites linked to CVE entries, the crawler efficiently gather large datasets of vulnerable code, further discussed in section \ref{irrelevantCommits}.

\subsection{IntraProject vs InterProject Vulnerabilities}
In the modern software development landscape, the complexity and scale of software systems have grown exponentially. With this growth, the challenge of managing software vulnerabilities has become more significant, especially when considering how vulnerabilities can propagate within and across different projects. The concepts of IntraProject and InterProject vulnerabilities are central to understanding the dynamics of how security flaws spread in software ecosystems.

IntraProject vulnerabilities often arise due to the reuse of code, modules, or libraries within the same project. For example, a vulnerability found in a core module of a project can affect multiple parts of the software that rely on this module. The primary concern here is that a single point of failure can compromise the entire system, especially in large-scale projects where a significant amount of the code is interdependent. This type of vulnerability is particularly prevalent in uniform software architectures, where a vulnerability in one part of the system can easily propagate to other parts due to the tightly coupled nature of the components. Whereas, InterProject vulnerabilities on the other hand, are security flaws that propagate from one project to another. This typically occurs in environments where software reuse is common, such as open-source ecosystems. Developers often incorporate external libraries, frameworks, or even entire codebases from other projects into their own, sometimes without fully understanding the security implications. The reuse of vulnerable code across different projects is a significant concern in open-source software development. For instance, a vulnerability in a widely used open-source library can affect thousands of other projects that depend on it. This can lead to widespread security incidents, as seen in high-profile cases like the Heartbleed bug in the OpenSSL library, which affected numerous software systems worldwide \cite{bojanova2023heartbleed}.

\subsection{SQVDT (Scalable Quantitative Vulnerability Detection Technique)}
SQVDT \cite{akram2021sqvdt} is a large-scale, automated source code vulnerability detection approach designed to identify known vulnerabilities by analyzing cloned code fragments in large repositories. It addresses the growing challenge of vulnerability propagation caused by code reuse across software systems. SQVDT works by extracting vulnerable code signatures from patch files linked to CVE (Common Vulnerabilities and Exposures) entries and identifying their occurrences in unpatched or reused code across massive open-source codebases, e.g., GitHub. The methodology involves three core components. First, a comprehensive vulnerability benchmark is constructed by mining vulnerability-related patch files from sources such as the National Vulnerability Database (NVD), GitHub repositories, and official vendor advisories. Second, a scalable code clone detection engine is applied to scan the source code from open-source GitHub projects. Finally, the system identifies vulnerable code instances and analyzes propagation patterns, determining whether vulnerabilities have spread within or across different projects.

In the results analysis, SQVDT demonstrated its practical effectiveness by identifying hundreds of previously undetected vulnerabilities in well-known software systems, including recent versions of the Linux kernel, Android variants like HTC and FindX kernels, and shared libraries commonly used in mobile and embedded systems. The evaluation of SQVDT was performed using standard metrics (e.g., True-Positive, False-Positive). SQVDT achieved an exceptionally low false positive rate of just 1\%, which reflect the tool's strong accuracy in identifying true vulnerable instances. The tool’s scalability was proven by its ability to analyze over seven TB of code efficiently using a Hadoop-based distributed processing environment. SQVDT offers a robust and scalable solution for vulnerability detection and propagation analysis in open-source ecosystems. It provides high accuracy, practical scalability, and fine-grained detection capabilities. Its design makes it suitable not only for academic research but also for industrial use in security auditing, vulnerability triage, software maintenance, and compliance monitoring across large and complex codebases.

\section{Experimental Setup and Results} \label{sec:Results}

\subsection{Analyzing a Patch File}
A vulnerability patch file is a specialized update designed to address and fix security vulnerabilities in a software program. These vulnerabilities can be weaknesses, flaws, or bugs that could be exploited by malicious actors to gain unauthorized access, cause disruptions, or steal sensitive information. The primary purpose of a vulnerability patch file is to close these security gaps and protect the software from potential attacks. During our analysis to vulnerabilities, we found that a vulnerability patch file contains different types of modifications to the existing code. These modifications can be the code changes (corrections or alterations to the source code to eliminate the vulnerability), configuration updates (configuration settings that may have contributed to the vulnerability), or security enhancements (implementation of new security measures). Vulnerability patch files are essential in maintaining the security and integrity of software systems.

\begin{figure}
  \centering
  \includegraphics[scale=.31]{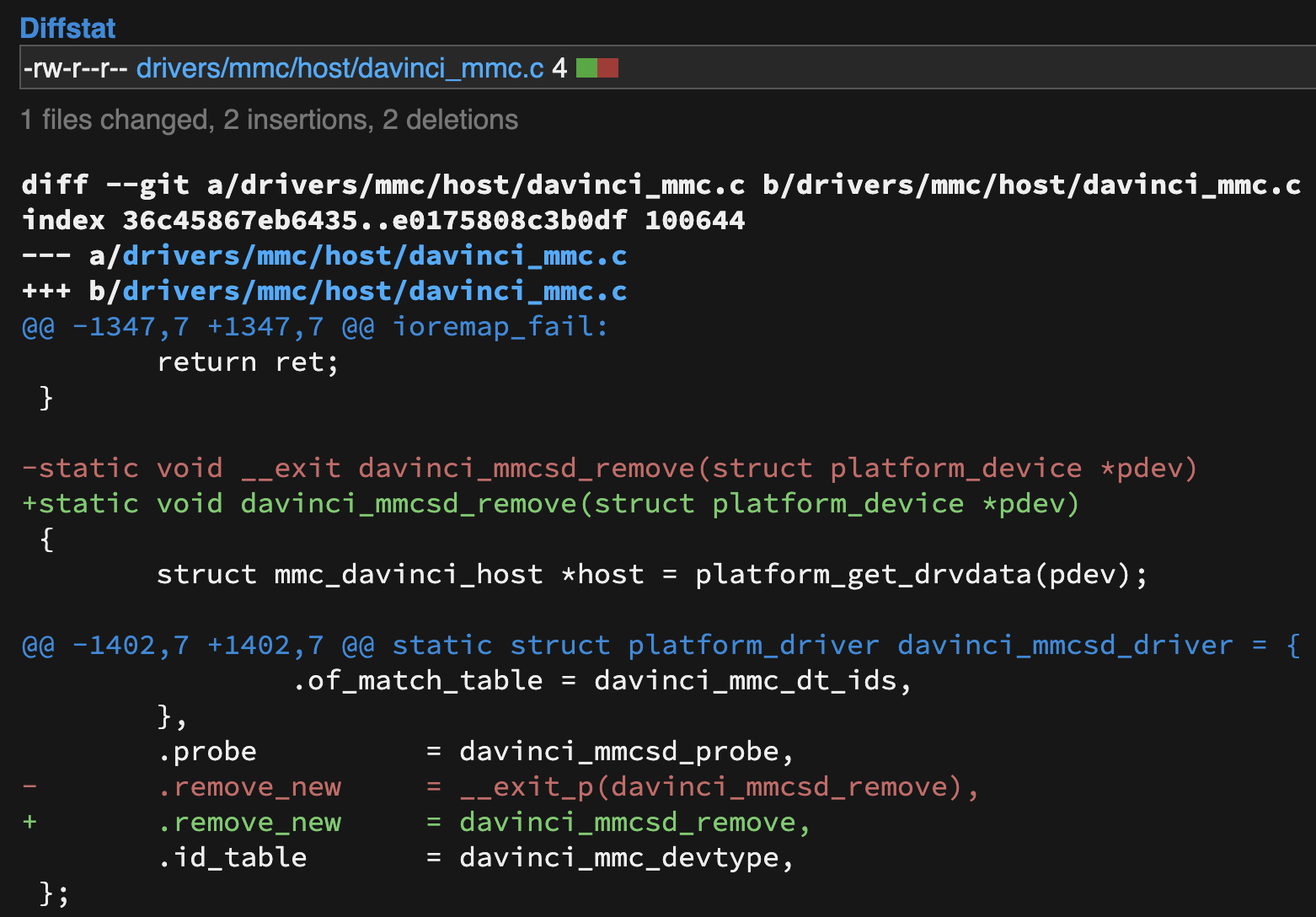}
  \caption{Patch file of CVE-2024-39484}
  \label{CVE-2024-39484}
\end{figure}
An example patch file of CVE-2024-39484 \footnote{https://nvd.nist.gov/vuln/detail/CVE-2024-39484} of 5.5 CVSS score (severity level) is shown in \figurename~\ref{CVE-2024-39484}, which has been released to fix "mmc: davinci: don't strip remove function when driver is builtin. \textit{Using \_exit for the remove function results in the remove callback being
  discarded with CONFIG\_MMC\_DAVINCI=y. When such a device gets unbound (e.g.
  using sysfs or hotplug), the driver is just removed without the cleanup
  being performed. This results in resource leaks. Fix it by compiling in the
remove callback unconditionally. This also fixes a W=1 modpost warning}". In \figurename ~\ref{CVE-2024-39484}, the `` + " symbol shows the addition in the code, the `` - " symbol shows the deletion in the code and `` @@ " shows the starting and ending lines of changes to make. There are changes made in this patch file at three different places. Furthermore, patch files also includes the address of the old file (before patch applied) and the new file (after patch applied), which are also shown in figure with \textbf{- - - a} and \textbf{+ + + b} respectively. So, our main concerned tag was \textbf{- - - a}, which is considered as a vulnerable file. We also extracted the particular function/method (e.g., \textit{davinici\_mmcsd\_remove()}) from this vulnerable file in which this changes were supposed to be made.

\subsection{Formation of Data Sets} \label{irrelevantCommits}
For experiments, we needed three kind of datasets; first CVE based vulnerability dataset, second open source software, third AI generated vulnerable code fragments. To collect the first two. We developed a web crawler to download vulnerable files from various web sources. The reference links related to different published vulnerabilities of some open source software were provided on the CVE MITRE platform\footnote{https://cve.mitre.org/} for reference purpose. Some of these references are purely informational, so we disregarded those that did not have \textit{commit} labels or downloadable patch files. We downloaded the entire list of approximately 112 K CVE entries from this site. During our exploration, we discovered that many CVE entries lack references to their patch files, which makes it challenging to obtain the vulnerable source code. Therefore, we focused on specific reference sites where the patch files were available and used them to locate the corresponding unpatched source code files.

To compile a comprehensive collection of vulnerable source code, we created various web crawlers tailored to different reference sites. A few references sites includes, kernel.org git repository\footnote{https://git.kernel.org/}, GitHub repository\footnote{https://github.com/}, Openssl repository\footnote{https://git.openssl.org}, etc.
Vulnerability dataset consists on hundreds of vulnerable source code files, which were extracted from a very reliable source (cve.mitre.org). This dataset consists on already publicly known cybersecurity vulnerabilities. In total we downloaded almost 3000 vulnerable files against their CVEs. A second dataset was open source projects, that were downloaded from GitHub in huge number of about 150K projects. These projects were random downloaded, consists of unique projects as well as different versions of the same projects. It was an old dataset, we used in one of our previous paper SQVDT \cite{akram2021sqvdt}, consists of 7-TB C/C++ source code (in which vulnerabilities were detected). Table 1 and Table 2, shows the downloaded dataset of open source projects of 7-TB C/C++ source code and the vulnerable code dataset against its CVE-number respectively.

\begin{table}
  \centering
  \caption{OSS dataset collection}
  \begin{adjustbox}{}
    \begin{tabular}{ccccc}
      \toprule
      Language & size & No of projects & No of Files & LOC \\
      \midrule
      C \& C++ & 7 TB & 152,823 & 235 (millions) & 83 (billions) \\
      \bottomrule
    \end{tabular}
  \end{adjustbox}
  \label{Dataset}
\end{table}

\begin{table}
  \centering
  \caption{Vulnerability dataset collection}
  \begin{adjustbox}{}
    \begin{tabular}{cccc}
      \toprule
      Source & No of CVE's & Number of Files & LOC   \\
      \midrule
      git.kernel.org  & 1,168 & 1,999 & 2,256,833 \\
      git.videolan.org & 194 & 202 & 227,802   \\
      cgit.freedesktop.org & 131 & 225 & 239,009   \\
      git.qemu.org & 123 & 138 & 189,458   \\
      git.savannah.gnu.org & 91 & 116 & 178,409   \\
      git.openssl.org & 85 & 130 & 149,565  \\
      git.ghostscript.com & 79 & 121 & 123,510   \\
      \midrule
      Total  & 1,871 & 2,931 & 3,364,586   \\
      \bottomrule
    \end{tabular}
  \end{adjustbox}
  \label{Dataset}
\end{table}

\subsection{Detection of Vulnerabilities Propagation}
For the detection of vulnerabilities in open source software, we used our vulnerability detection tool \textit{SQVDT (Scalable Quantitative Vulnerability Detection Technique)} \cite{akram2021sqvdt}. This tool is specifically designed to address the challenges of large-scale vulnerability detection by providing a robust and scalable approach to analyzing source code for potential security weaknesses. SQVDT  detects vulnerabilities by employing a code clone and quantitative analysis of source code. The methodology involves parsing the source code to create a a series of tokens of the code, followed by the algorithms to extract different features from the code. These features then used to identify potential vulnerabilities in the code, including both known and previously unrecognized vulnerabilities. The tool scales efficiently across large codebases by breaking down the analysis into smaller, manageable segments, ensuring comprehensive coverage without compromising performance. By automating this process, SQVDT can rapidly and accurately detect vulnerabilities that might be missed during manual reviews, making it a powerful tool for enhancing software security. Our study involved running this tool on a diverse set of open-source projects to assess their vulnerability profiles.

The experiments conducted with \textit{SQVDT} revealed several critical vulnerabilities across multiple projects, demonstrating the tool's capability to uncover security issues that might go unnoticed during manual code reviews or traditional testing methods. By automating the vulnerability detection process, the tool not only accelerates the identification of security risks but also improves the overall reliability of the software by reducing the chances of exploitable flaws being deployed in production environments. The tool's quantitative approach ensures that even subtle vulnerabilities, which might be overlooked due to their complexity, should be detected. More than 12 thousands vulnerable files were detected in almost 115 thousands open source projects. Most of the detected vulnerable files were the propagated vulnerabilities from other systems. A precise set of results (detected vulnerabilities) are discussed in this paper for reference to vulnerability propagation.

\begin{figure}
  \centering
  \includegraphics[scale=.22]{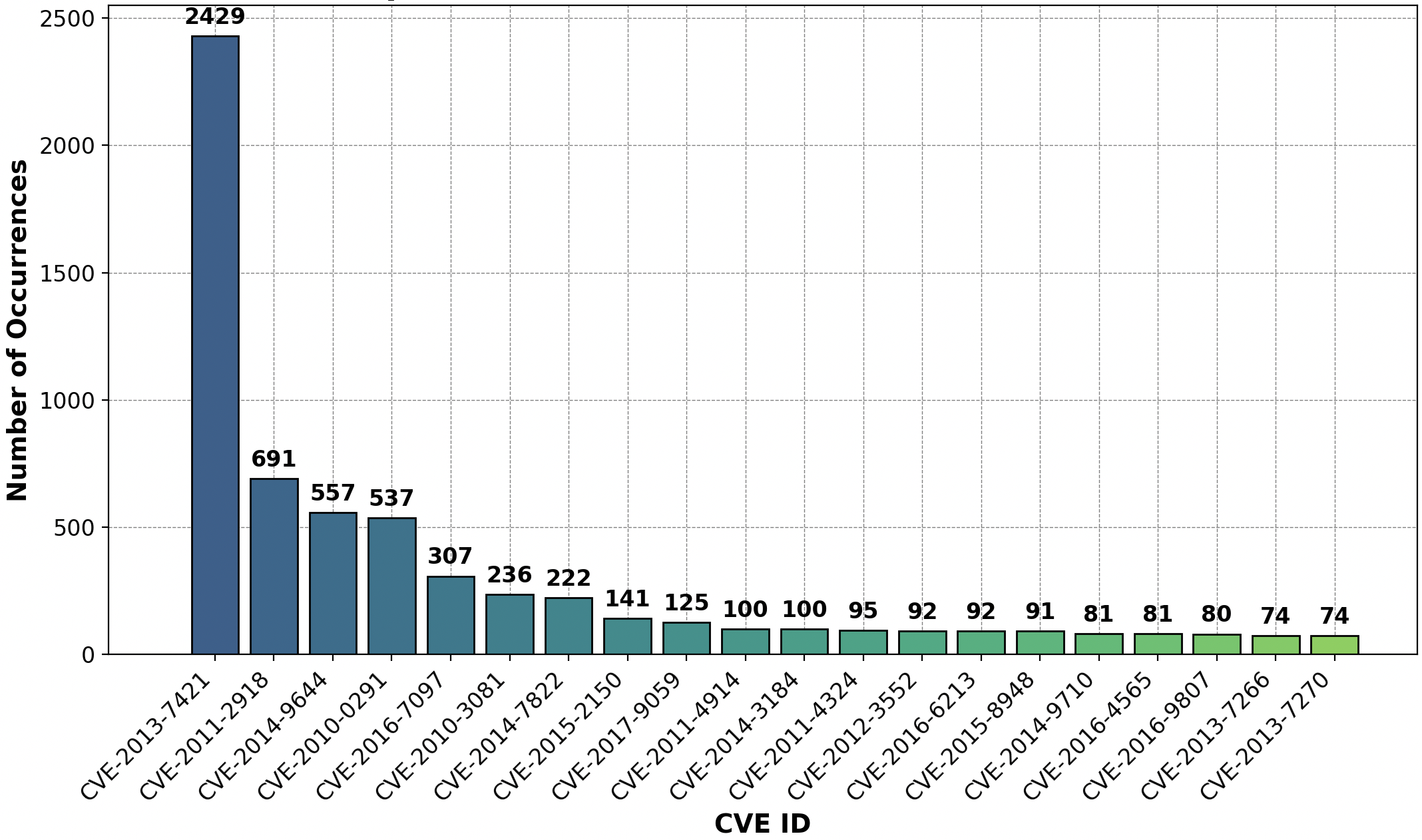}
  \caption{Top 20 most propagated CVE based vulnerabilities}
  \label{topCVE}
\end{figure}

\subsection{Most Propagated Vulnerabilities (CVEs)} \label{sec:mostPropagated}
\begin{figure}
  \centering
  \includegraphics[scale=.35]{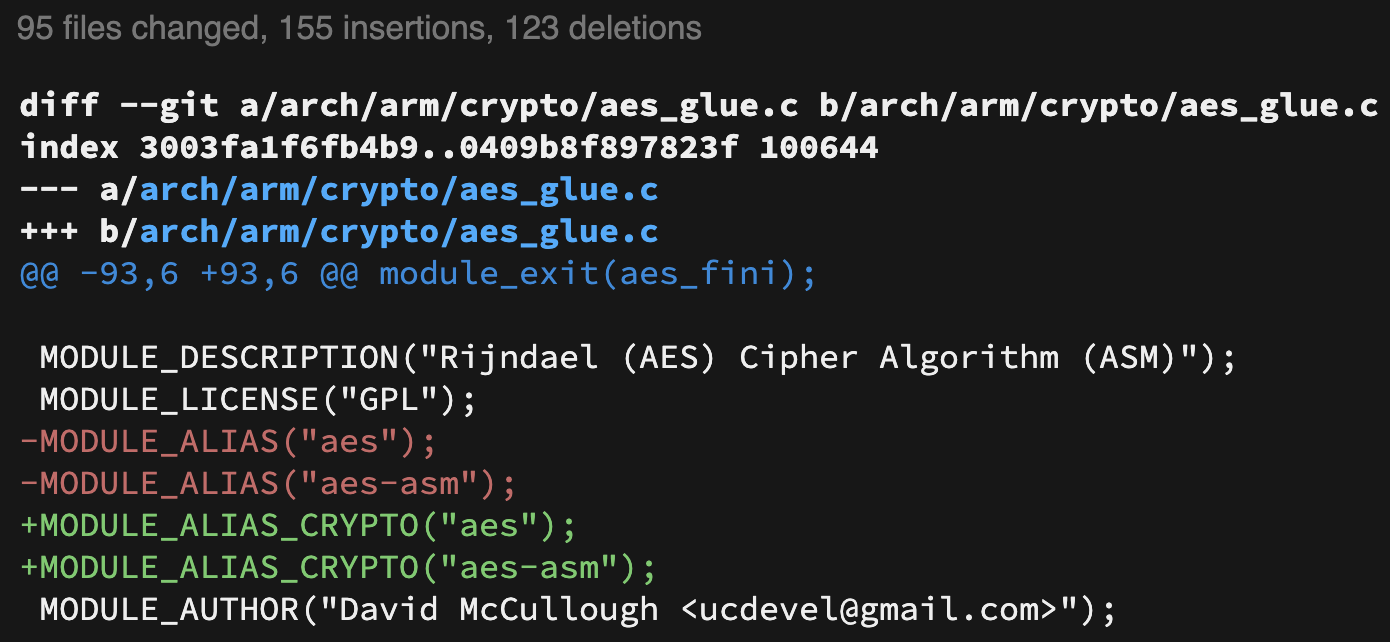}
  \caption{Patch information of CVE-2013-7421}
  \label{2013-7421}
\end{figure}

Lets discuss most often propagated vulnerabilities that were detected in several projects available on GitHub that shares the same vulnerable file or a code fragment of that vulnerable file. Despite the availability of published patch files (vulnerability fixes), these projects still remained unpatched, and users continue to utilize them unknowingly. Top 20 most propagated CVE based vulnerabilities that are discovered in various open-source projects are shown in \figurename~\ref{topCVE}. Each bar represents the number of times a particular CVE has been identified across different projects, indicating its level of propagation. It emphasizes the importance of identifying and addressing highly propagated vulnerabilities to enhance the security posture of open-source systems. This figure highlights key areas where security efforts should be concentrated to prevent widespread exploitation. CVEs with a high number of occurrences, such as \textit{CVE-2013-7421}, indicate a vulnerability that have propagated extensively across different projects. This could be due to common dependencies or components used in a wide array of systems, making them more prone to exploitation and harder to mitigate on a large scale. The graph suggests that efforts to secure open-source systems should prioritize these highly propagated vulnerabilities, as they pose a significant risk to a larger number of systems. The presence of older CVEs, like those from 2010 and 2011, in this top 20 list implies that many systems still have not fully patched yet. Let's discuss the root cause of the three most occurred vulnerabilities that propagated to other systems.  \\
\textbf{CVE-2013-7421:} This vulnerability was detected in our open source dataset for almost 2429 times. To understand the reason why it was propagated to other systems too many time, let's discuss the nature of this vulnerability. This vulnerability first time discovered in 2013 in Linux kernel. \textit{It exists in the Crypto API of Linux kernel before version 3.18.5, which allows local users to load an arbitrary kernel modules via a bind system call for an AF\_ALG socket with a module name in the salg\_name field} \footnote{https://nvd.nist.gov/vuln/detail/CVE-2013-7421}. If we look at the information regarding its patch file as shown in \figurename~\ref{2013-7421}, from the first line it shows that the patch was applied on 95 files with 155 insertions and 123 deletion, one of the change (in one file) is shown in the figure. Because the changes were applied on several files, and those vulnerable files were part of our vulnerability data set. Therefore, we might have retrieved several vulnerable open source projects against this CVE, which shows that several projects might have shared the similar unpatched files in their code. The propagation of this vulnerability across numerous open-source projects underscores the importance of rigorous dependency management and the need for prompt patching to lower the security risks in the shared software components.\\

\begin{figure}
  \centering
  \includegraphics[scale=.23]{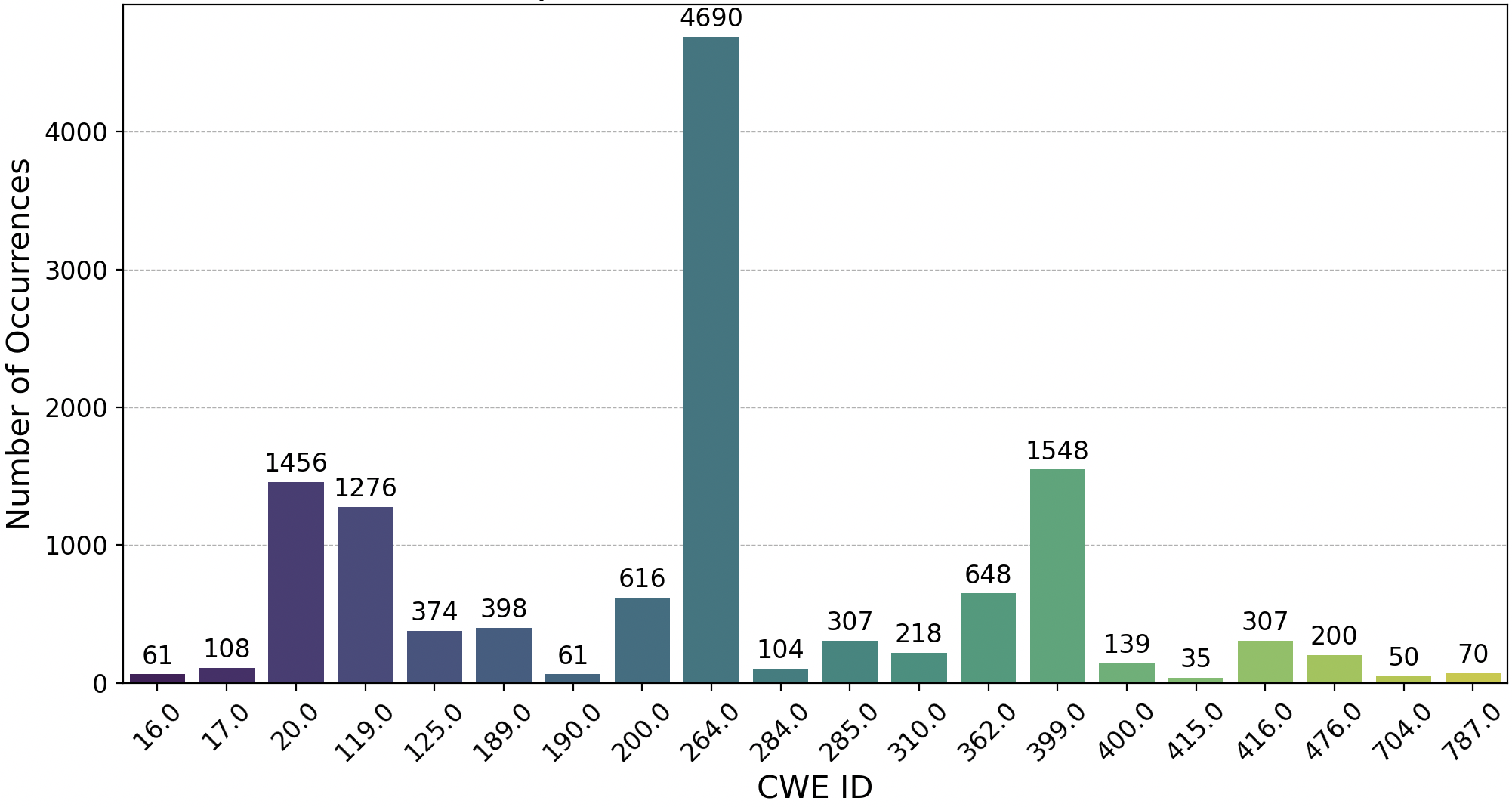}
  \caption{Top 20 CWE-ID (Common Weakness Enumeration) propagation }
  \label{top20cwe}
\end{figure}

\textbf{CVE-2011-2918: } This vulnerability was announced in 2011 in Linux operating system. This vulnerability is about the\textit{ performance events of subsystem in the Linux kernel before version 3.1, that does not properly handle event overflows associated with PERF\_COUNT\_SW\_CPU\_CLOCK events, which allows local users to cause a denial of service (system hang) via a crafted application} \footnote{https://nvd.nist.gov/vuln/detail/CVE-2011-2918}. The patch of this CVE was applied on 46 different files with 119 additions and 141 deletions. The vulnerability occurs due to a lack of proper checks on certain operations within the RDS protocol, allowing an attacker to execute code that could crash the system or allow a root access. This makes it a particularly dangerous vulnerability, especially in multi-user environments where untrusted users might have access to a shared system. The propagation of this vulnerability seems to be more in various Linux distributions and largely due to the widespread adoption of the kernel and its components. Because the Linux kernel is at the heart of many open-source and commercial systems, any vulnerability within it can have far-reaching consequences. Systems that did not update their kernel or apply security patches in a timely manner were especially vulnerable to attacks exploiting this vulnerability. This widespread impact underscores the importance of continuous monitoring, patching, and updating of critical software components, especially in a widely deployed and shared environments like those which is powered by the Linux kernel.\\
\textbf{CVE-2014-9644: }This vulnerability is is also related to Linux, \textit{the Crypto API in the Linux kernel before version 3.18.5 allows local users to load arbitrary kernel modules via a bind system call for an AF\_ALG socket with a parenthesized module template expression in the salg\_name field, as demonstrated by the vfat(aes) expression} \footnote{https://nvd.nist.gov/vuln/detail/CVE-2014-9644}. The patch of this vulnerability updated 23 files, with 26 insertions, and 2 deletions. This vulnerability propagates extensively due to its presence in widely used software or libraries. Its propagation is accelerated by the reuse of vulnerable components in different projects, including those relying on third-party libraries. The widespread adoption of the affected software and insufficient patch management further contribute to the persistence and spread of this vulnerability across multiple platforms and projects.

\subsection{Common Weakness Enumeration Propagation} \label{sec:CWE}
The Common Weakness Enumeration (CWE) is a category of a system for hardware and software weaknesses and vulnerabilities. It has a standardized identifier (ID) used to classify software security weaknesses or vulnerabilities. The CWE system categorizes and catalogs these weaknesses, allowing developers, researchers, and security professionals to identify and address common software security issues. Each CWE-ID corresponds to a specific type of vulnerability, such as \textit{improper input validation, buffer overflows, or improper access controls, etc}. The CWE list is widely used in cybersecurity to improve software quality and enhance security practices. It is sustained by a community project with the goals of understanding flaws in software and hardware and creating automated tools that can be used to identify and fix these flaws.

\figurename~\ref{top20cwe} illustrates the frequency of the most commonly propagated CWEs within different projects. The presence of these CWEs highlights systemic issues in handling \textit{access controls, resource management, input validation, and memory management}, which are crucial areas to address for improving software security across projects. On the top of the list \textit{CWE-264 (Permissions, Privileges, and Access Controls)}\footnote{https://cwe.mitre.org/data/definitions/264.html} dominates with 4690 occurrences, reflecting the widespread challenge in managing access rights across different projects. This high number suggests that access control issues are a significant factor in the vulnerability propagation. \textit{CWE-399 (Resource Management Errors)}\footnote{https://cwe.mitre.org/data/definitions/399.html} ranks second with 1548 occurrences, emphasizing the criticality of efficient resource management in avoiding vulnerabilities. These errors often arise when software fails to properly manage system resources such as memory, file handles, or network connections, leading to potential exploits.

\textit{CWE-20 (Improper Input Validation)}\footnote{https://cwe.mitre.org/data/definitions/20.html} and \textit{CWE-119 (Improper Restriction of Operations within the Bounds of a Memory Buffer)}\footnote{https://cwe.mitre.org/data/definitions/119.html} also show significant frequencies, with 1456 and 1276 occurrences, respectively. These issues typically stem from failing to properly validate user inputs and improperly managing memory, which can lead to severe vulnerabilities, including buffer overflows. Other noteworthy CWE in the top 20 propagated CWEs include \textit{CWE-362 (Concurrent Execution using Shared Resource with Improper Synchronization)} with 648 occurrences, \textit{ CWE-284 (Improper Access Control)} with 104 occurrences, and \textit{CWE-476 (NULL Pointer Dereference)} with 200 occurrences. These CWEs further indicate common areas of concern in software security, particularly in managing concurrency, access controls, and memory references.

\subsection{Critical (High Risk) Propagated Vulnerabilities} \label{sec:critical}
High risk vulnerabilities are security vulnerabilities that pose a significant threat to the systems and require immediate attention. They are classified based on the Common Vulnerability Scoring System (CVSS), which assigns a severity score from 0 to 10. A CVSS score between 8 and 10 indicates criticality \cite{aggarwal2023study} of the vulnerability. If we encounter such vulnerabilities, it's crucial to address them promptly to low down the potential risks. CVSS is a published standard used by organizations worldwide.

\begin{figure}
  \centering
  \includegraphics[scale=.24]{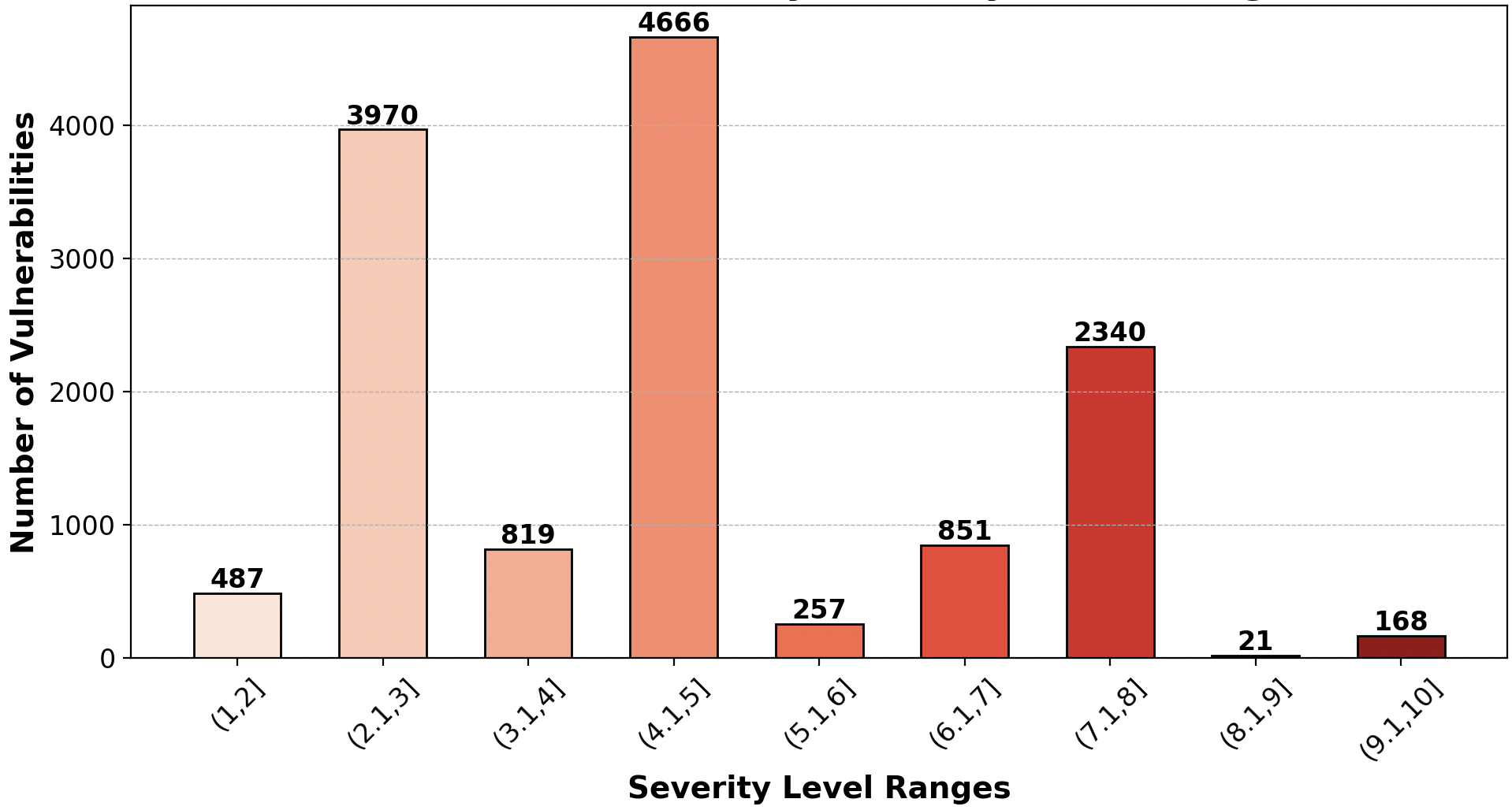}
  \caption{Number of propagated vulnerabilities by their severity level (CVSS: 1-10)}
  \label{severityLevel}
\end{figure}

In our detected vulnerabilities, we categories them according to their CVSS score, for the better analysis of vulnerability propagation with respect to their severity level as shown in \figurename~\ref{severityLevel}. The figure provides a detailed view of how vulnerabilities spread across different severity levels in open-source projects. This analysis is crucial for understanding the risks associated with open-source software, which often forms the backbone of many applications and systems. Let's discuss few most propagated severity levels along with the open source projects.\\
\textbf{Severity Levels [4.1 to 5]:} This range has the highest count with 4666 vulnerabilities, indicating that many propagated vulnerabilities fall into medium severity. This suggests that many vulnerabilities, while not immediately critical, pose a significant risk if left unaddressed. Medium severity vulnerabilities are often pervasive because they might be more challenging to detect and remove, especially in complex systems with multiple dependencies. The large number of vulnerabilities in this range indicates that many open-source projects might be exposed to moderate risks that could be exploited under certain conditions.\\
\textbf{Severity Levels [2.1 to 3]:} The second highest propagation detected is 3970 vulnerabilities, suggesting a substantial number of low-to-medium severity issues. These vulnerabilities are often overlooked because they may not cause immediate or severe damage. However, when accumulated, even low-severity vulnerabilities can weaken the overall security posture of a project, making it easier for attackers to exploit them in combination with other vulnerabilities. This large count suggests that many open-source projects may not be thoroughly addressing these lower-severity issues.\\
\textbf{Severity Levels [7.1 to 8]:} High-severity vulnerabilities are significant, with 2340 occurrences, making them a critical concern. These vulnerabilities are critical because they can lead to severe consequences, such as data breaches, system takeovers, or significant service disruptions. The fact that there are 2340 vulnerabilities in this range highlights the need for immediate attention and remediation, especially in open-source projects where updates and patches might not be as frequent or controlled as in proprietary software.\\
\textbf{Severity Levels [8 to 10]:} While the most severe (high risk) vulnerabilities are fewer (189), they represent the most critical threats and are a focus for security efforts. While the count is lower compared to other ranges but these vulnerabilities are the most dangerous and demand urgent remediation. The lower number is expected since vulnerabilities of this magnitude are usually rare but can have severe implications, potentially affecting thousands of systems worldwide.

\begin{figure}
  \centering
  \includegraphics[scale=.24]{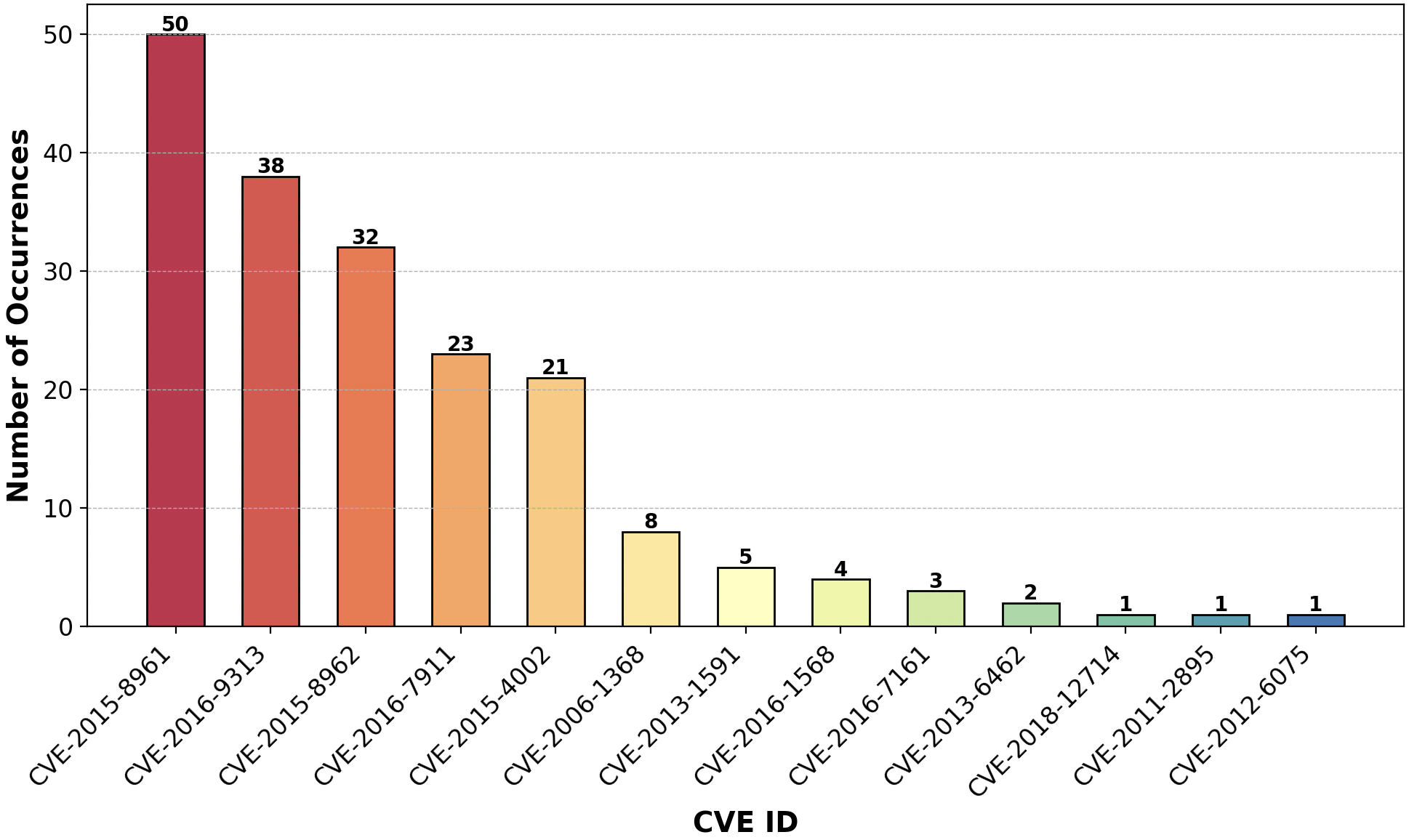}
  \caption{Critical propagation detected of range (CVSS: 9-10)}
  \label{severeCVE}
\end{figure}

\begin{figure}
  \centering
  \includegraphics[scale=.24]{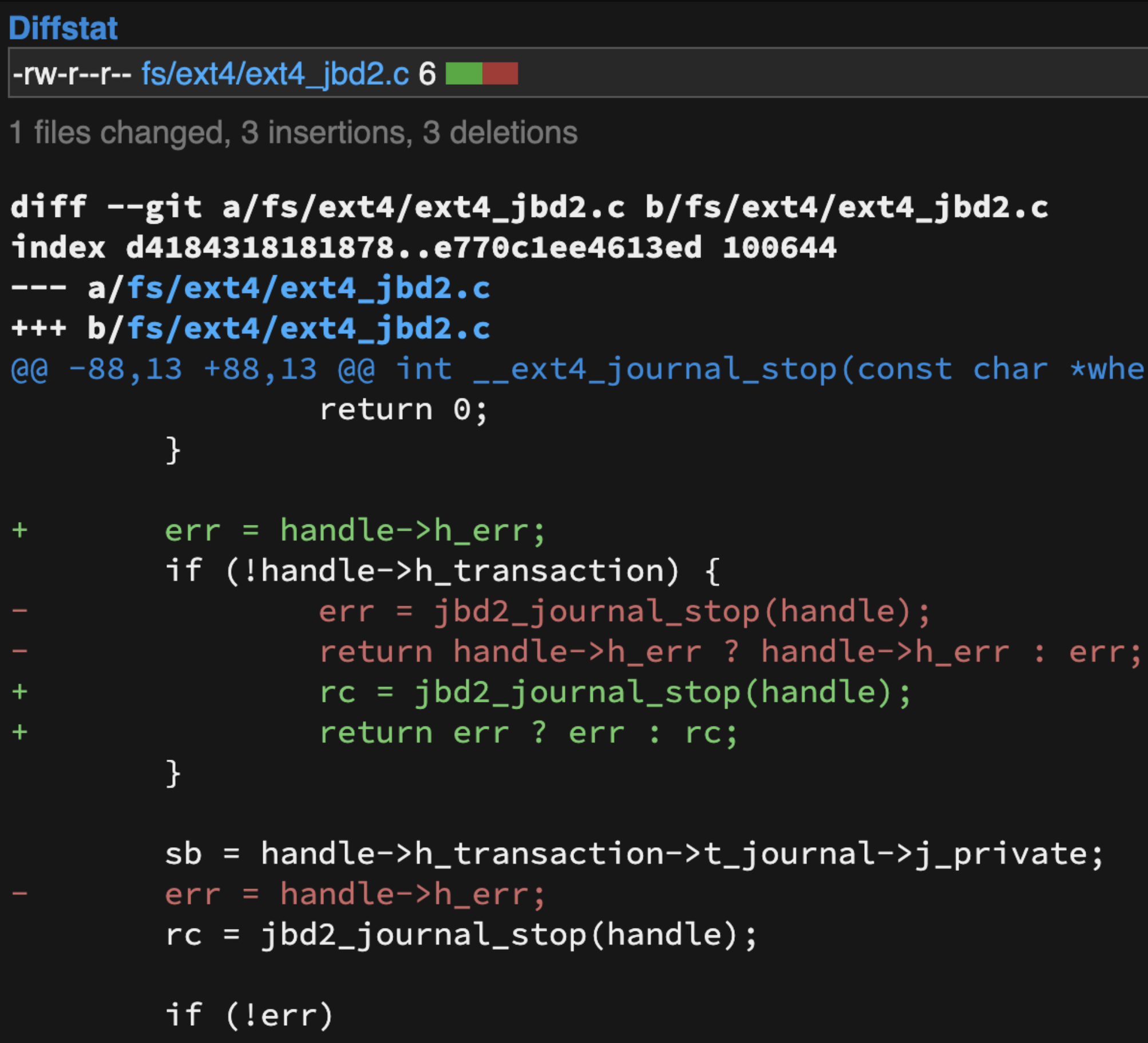}
  \caption{CVE-2015-8961 patch file in the Linux-kernel}
  \label{CVE-2015-8961}
\end{figure}

\figurename~\ref{severeCVE} shows the distribution of high risk vulnerabilities (189), detected by (SQVDT) in open source systems of CVSS score [8-10]. From the figure, we can see that \textit{CVE-2015-8961} \footnote{https://nvd.nist.gov/vuln/detail/CVE-2015-8961} leads the chart with 50 occurrences, which mean that this high risk vulnerability was detected in several projects about 50 times. This CVE is likely a critical vulnerability that has been widely propagated across various projects, making it a significant security concern. This vulnerability was first time announced in 2015 in Linux-kernel before version 4.3.3 and it has a very critical CVSS score of 9.3. A patch (vulnerability fix) file of this vulnerability is shown in \figurename~\ref{CVE-2015-8961}. By exploring a patch file of this vulnerability, we can see that a function in the file \textit{fs/ext4/ext4\_jbd2.c} allows a local users to gain fully privileges and cause a denial of service attack by leveraging an improper access to a specific error field.

Another vulnerability, \textit{CVE-2016-9313} follows with 38 occurrences in the results, indicating another critical vulnerability with extensive propagation. While \textit{CVE-2015-8962} and \textit{CVE-2016-7911} also show a significant propagation with 32 and 23 occurrences, respectively. Whereas, the remaining vulnerabilities show a decreasing trend in propagation, with some vulnerabilities like \textit{CVE-2018-12714} and \textit{CVE-2011-2895} having only 1 occurrence. This distribution highlights that while a few vulnerabilities are highly propagated and represent a substantial risk, others have limited propagation but can still pose critical threats to individual systems. The variation in propagation levels could be due to differences in the adoption of affected software components, the ease of exploitability, or the nature of the vulnerability itself. \figurename~\ref{severeCVE} underscores the need for targeted security efforts on the most propagated and critical vulnerabilities to lower down the widespread risks effectively. This distribution underscores the need for robust vulnerability management practices in open-source projects, especially focusing on the critical [7, 10] severity levels, to prevent widespread exploitation and ensure the security of dependent systems.

\subsection{IntraProject Propagated Vulnerabilities}
These security flaws spread within a single software project, maybe in between different versions of the same project. These vulnerabilities often arise from code reuse, where the same piece of vulnerable code is used at multiple places within a project. This can occur when developers copy-paste code across different modules or functions without realizing that the code contains a vulnerability. As the code propagates within the project, it becomes more challenging to identify and fix the issue, potentially leading to widespread security risks within that project. There were three main project categories in our detected results, i.e., \textit{Linux}, \textit{Android}, \textit{Qemu}. Therefore, we extracted these most prominent projects from our results, and consider them as a case study of IntraProject vulnerability propagation analysis. We explore the root causes of how these vulnerabilities were propagated in between these projects. Let's discuss these case studies one by one in more detail.

\begin{figure}
  \centering
  \includegraphics[scale=.24]{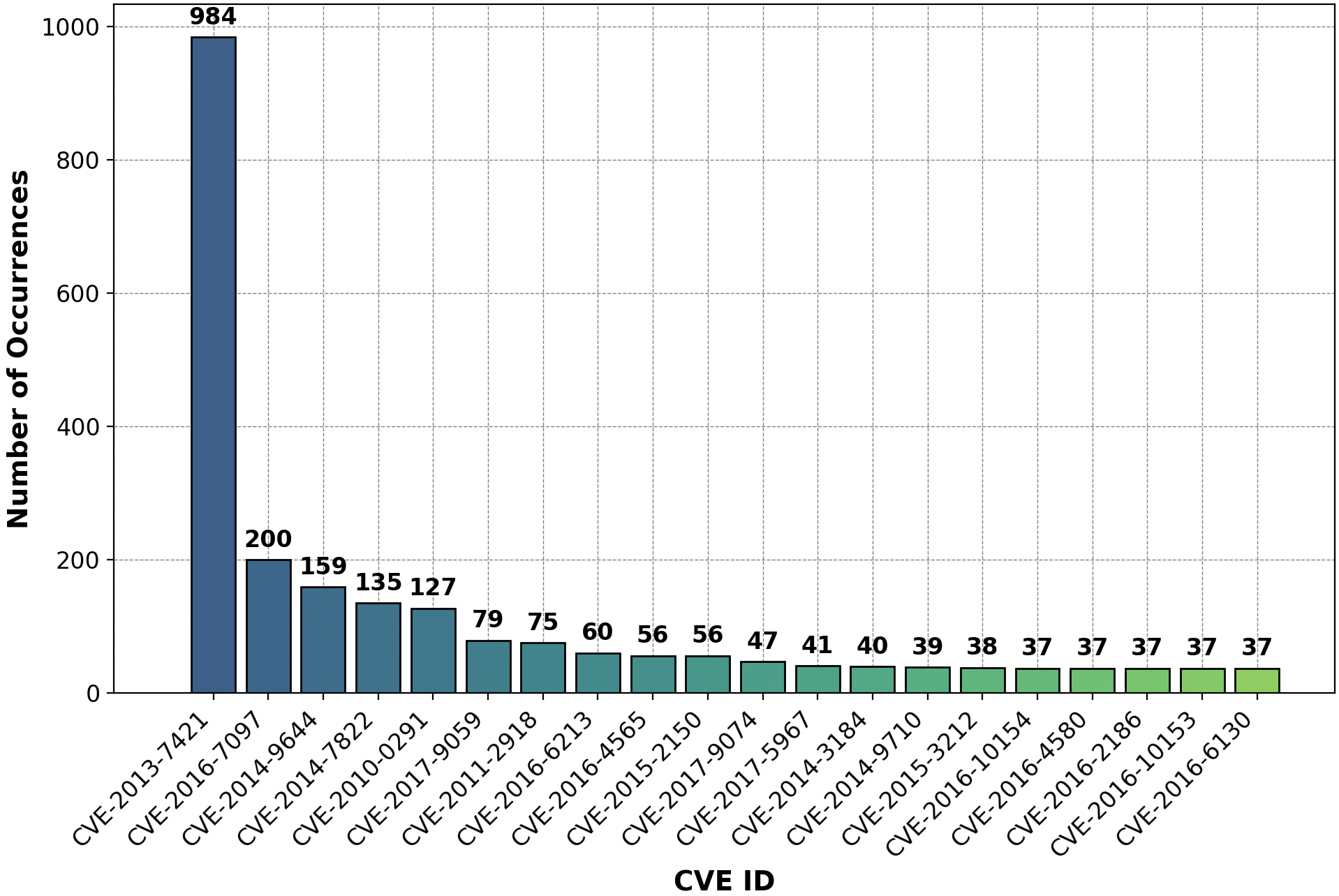}
  \caption{Top 20 most IntraProject vulnerabilities in Linux }
  \label{linux20}
\end{figure}

\subsubsection{Vulnerabilities in Linux}
IntraProject propagated vulnerabilities in Linux arise when a single vulnerability within the Linux codebase is inadvertently spread across multiple components or modules of the same project. \figurename~\ref{linux20} shows
the top twenty most IntraProject propagated vulnerabilities in Linux operating system. The most propagated vulnerability is \textit{CVE-2013-7421}\footnote{https://nvd.nist.gov/vuln/detail/CVE-2013-7421}, which appears 984 times, indicating it is a highly impactful issue within various components of the Linux system. \textit{CVE-2016-7097} follows with 200 occurrences, and other notable CVEs like \textit{CVE-2014-9644} and \textit{CVE-2014-7822} are also frequently propagated, showing 159 and 135 occurrences respectively. This distribution highlights how certain vulnerabilities have significant reach within Linux, potentially affecting multiple subsystems or software components, leading to widespread security risks. This propagation often occurs due to the reuse of vulnerable code segments, complex inter-dependencies between different modules. In Linux, where the codebase is vast and modular, these vulnerabilities spread quickly, and affects various parts of the system. The consequences can be severe, as the propagation may lead to widespread security risks, making it difficult to isolate and patch the vulnerability effectively. One major reason for this is the extensive use of shared libraries, modules, and the Linux kernel in between different versions of Linux. When a vulnerability exists in a commonly used library, it can propagate to all the applications and components that depends on it. The modular nature of Linux, while beneficial for flexibility and customization, also creates opportunities for vulnerabilities to be unintentionally propagated during updates, or maybe when new features are integrated.

Moreover, legacy code within Linux, often maintained for compatibility reasons, can harbor vulnerabilities that are accidentally introduced into newer code. Developers might copy and paste code snippets from older modules, unintentionally spreading these vulnerabilities. This problem is compounded by the fact that the Linux kernel and its associated modules are continually evolving, with contributions from a global community of developers. While this collaborative approach accelerates innovation, it also increases the likelihood of vulnerabilities spreading within the project. Another reason of this propagation can be that some people downloaded the Linux code and upload on GitHub in their own repository by modifying the code. Because we collected open source projects randomly, therefore, our code data set have this kind of Linux versions. The propagation of these vulnerabilities can have far-reaching effects, compromising the security of not just the Linux kernel but also the numerous systems and devices that rely on it, which would be discussed in InterProject propagation section \ref{sec:interSection} in more detail. To mitigate these risks, the Linux community must prioritize rigorous code audits, automated vulnerability scanning, and proactive patch management to detect and address propagated vulnerabilities before they can be exploited.

\begin{figure}
  \centering
  \includegraphics[scale=.25]{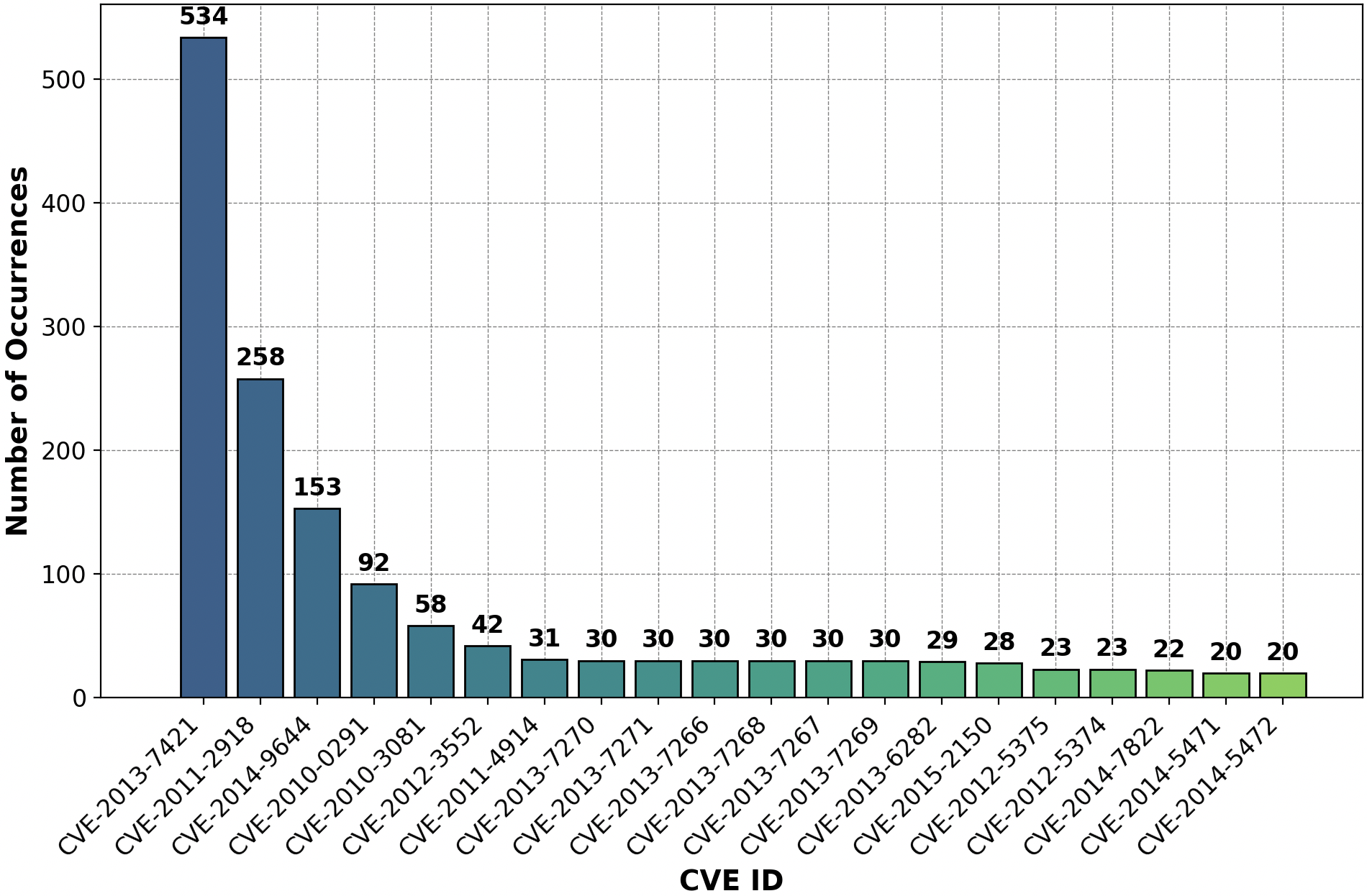}
  \caption{Top 20 most IntraProject propagated vulnerabilities in Android systems }
  \label{android20}
\end{figure}

\subsubsection{Vulnerabilities in Android}
IntraProject propagated vulnerabilities in Android represent a critical security challenge due to the operating system's vast and intricate architecture. \figurename~\ref{android20} shows most IntraProject propagated vulnerabilities in Android systems. The most propagated vulnerability in between Android systems is \textit{CVE-2013-7421} with the occurrence of 534. This vulnerability is the same as of Linux, which was announced in year 2013 in the Linux operating system. This shows that there is a possibility that this vulnerability in Androids came from Linux, as most of the Android operating systems use Linux-kernel as a backbone of their code. \textit{CVE-2011-2918}\footnote{https://nvd.nist.gov/vuln/detail/cve-2011-2918} is a very old Linux based vulnerability (with occurrence 258) of year 2011, that was also detected in the Android open source data set. It shows that some of the Android operating systems available on GitHub are not patched or maybe very old versions, which should be removed from the repository. Here another important concern is that our open source dataset consist of Android systems from different vendors, including \textit{Samsung, Huawei, Oppo, Xiaomi, etc}. On the analysis of all the propagated vulnerabilities detected in these Android systems that are shown in \figurename~\ref{android20}, we discovered that most of the vulnerabilities are Linux system vulnerabilities. IntraProject propagated vulnerabilities in Android are essentially inherited from its Linux-based kernel. Since Android is built on top of the Linux kernel, many of the vulnerabilities detected in Android systems originate from the underlying Linux codebase. These Linux vulnerabilities propagate within the Android ecosystem due to the shared components and libraries that both platforms utilize.

Android, being an open-source platform with numerous versions and customizations, is prone to vulnerabilities spreading across different components within the same project. These vulnerabilities often arise from the reuse of insecure code, shared libraries, or outdated dependencies. For instance, a security flaw in a core Android library could be inherited by various modules or even different versions of the Android OS by different vendors, leading to widespread exposure. This propagation is magnified by the extensive ecosystem of Android devices, which often lack in receiving security patches on time, that leaving many devices vulnerable. Addressing these vulnerabilities requires a proactive approach by all android vendors, including regular code audits, timely patching, and rigorous testing across all levels of the Android framework to prevent the spread of vulnerabilities.

\begin{figure}
  \centering
  \includegraphics[scale=.26]{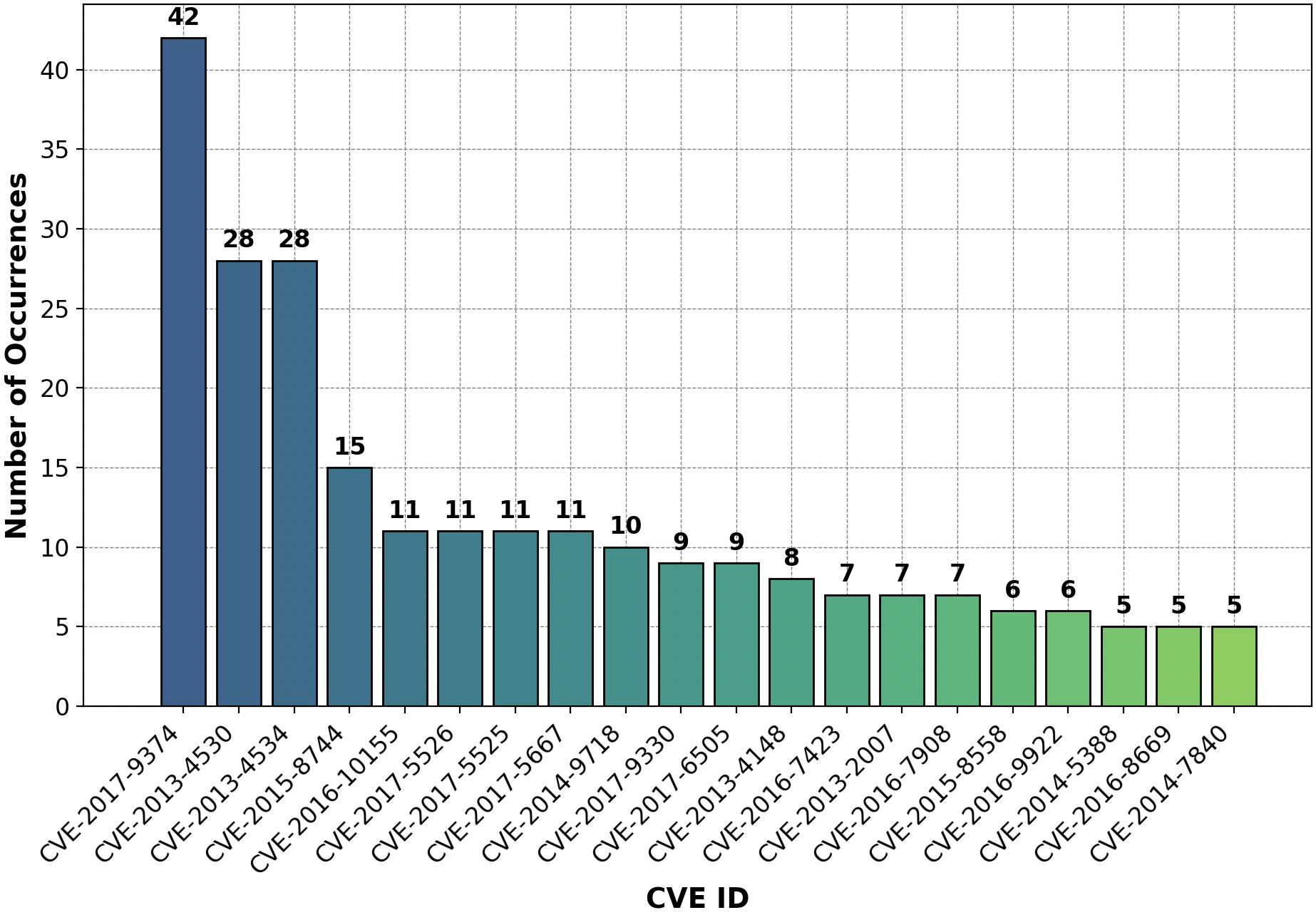}
  \caption{Top 20 most IntraProject propagated vulnerabilities in Qemu systems }
  \label{qemu20}
\end{figure}

\subsubsection{Vulnerabilities in QEMU}
Qemu \footnote{https://www.qemu.org/} was also a widely seen result in our detected vulnerabilities, therefore we also consider it as another case study for IntraProject vulnerability propagation analysis. Qemu being a widely used open-source emulator and virtualizer, has its own share of vulnerabilities. Vulnerabilities propagation in Qemu represent a critical concern due to the nature of Qemu as a versatile open source virtualization platform. These vulnerabilities often originate from shared or reused code within different components of the same project, making it possible for a single flaw to propagate across various modules. Since Qemu is used in a variety of environments, from desktop virtualization to cloud computing, these vulnerabilities can affect multiple systems, making them particularly dangerous. Additionally, because Qemu emulates hardware, vulnerabilities within it can have direct consequences on the virtualized operating systems. Addressing these vulnerabilities requires diligent tracking, testing, and patching across all instances where the vulnerable code might be deployed.

\figurename~\ref{qemu20} shows twenty most appeared IntraProject propagated vulnerabilities in Qemu, with highest appeared vulnerability \textit{CVE-2017-9374}, occurrence 42 times. This vulnerability is a \textit{memory leak in Qemu (aka Quick Emulator), when built with USB EHCI Emulation support, allows local guest OS privileged users to cause a denial of service (memory consumption) by repeatedly hot-unplugging the device} \footnote{https://nvd.nist.gov/vuln/detail/CVE-2017-9374}. \textit{CVE-2013-4530} and \textit{CVE-2013-4534} are also significant examples of vulnerabilities that propagated extensively within projects, primarily due to the widespread use of shared or similar codebases across multiple components. These both vulnerabilities found in Qemu versions before 1.7.2, that \textit{allows remote attackers to cause a denial of service or possibly execute arbitrary code via crafted tx\_fifo\_head and rx\_fifo\_head values}. The extensive propagation of these vulnerabilities highlights the critical importance of thorough testing and code review practices, especially in large and complex projects like Qemu, where a single vulnerability can have widespread implications.

\begin{figure}
  \centering
  \includegraphics[scale=.25]{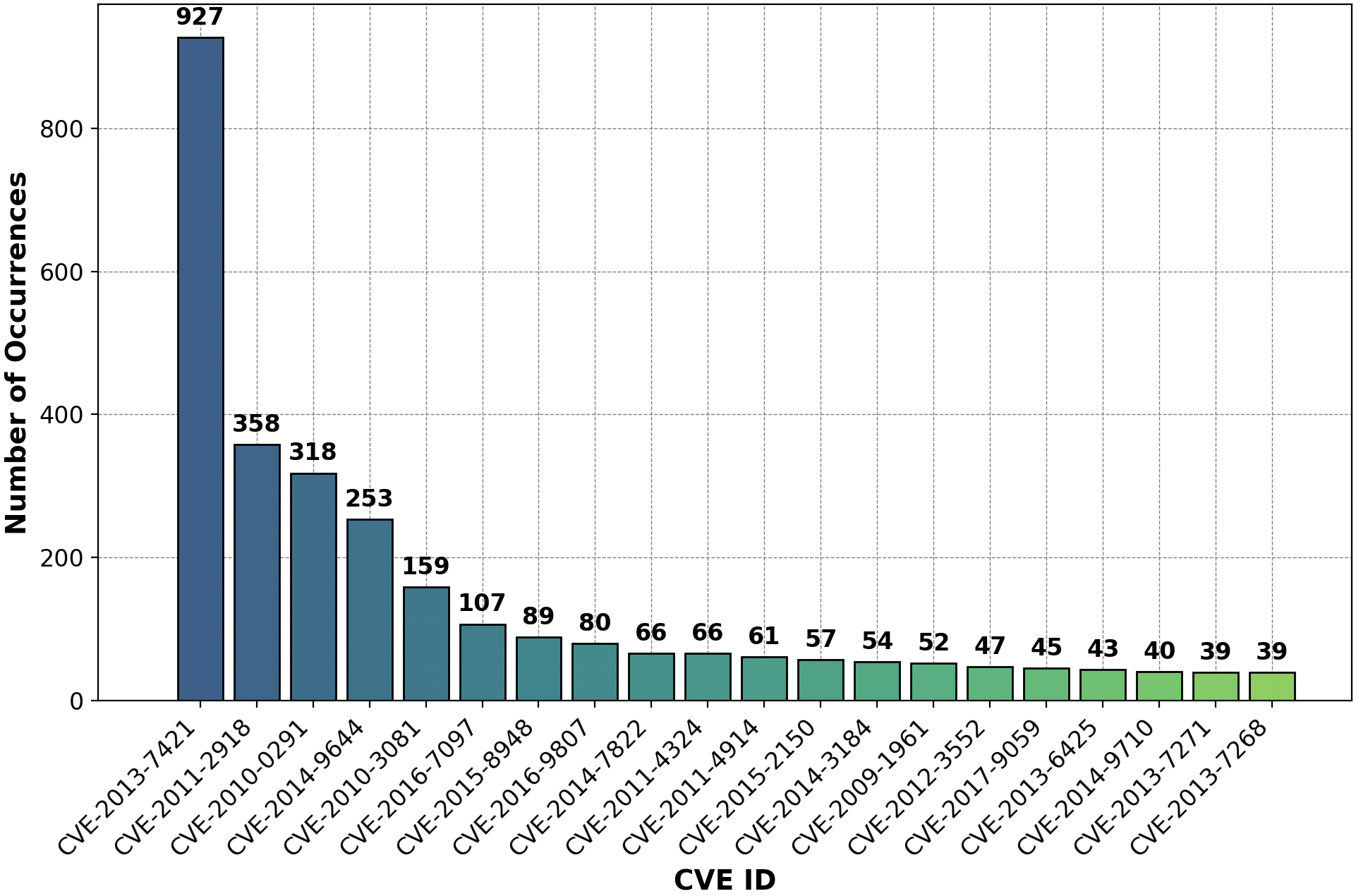}
  \caption{Top 20 most InterProject propagated vulnerabilities in different systems}
  \label{interProject20}
\end{figure}

\begin{figure}
  \centering
  \includegraphics[scale=.27]{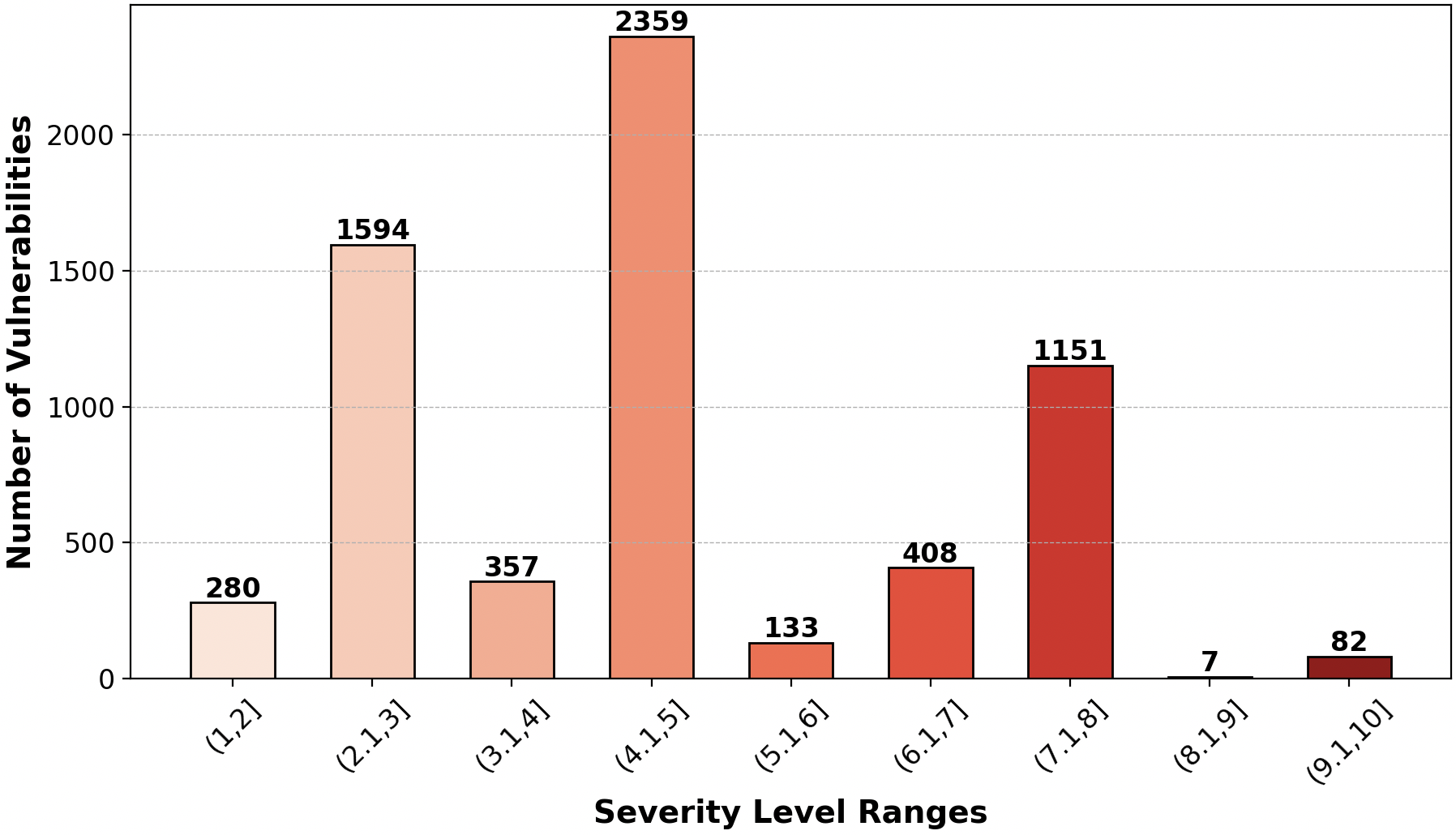}
  \caption{InterProject propagated vulnerabilities by their severity level (CVSS: 1-10)}
  \label{interprojectSeverity}
\end{figure}

\subsection{InterProject Propagated Vulnerabilities}\label{sec:interSection}
InterProject vulnerabilities are a significant concern, as vulnerabilities in one project can impact multiple other projects due to code reuse, shared libraries, or dependencies. In our analysis, we detected several instances where vulnerabilities originally found in one project were propagated to other projects, potentially affecting a wide range of systems. For instance, vulnerabilities such as \textit{CVE-2013-7421} (shown in \figurename~\ref{interProject20}) was initially identified in one environment but later found that it has been spread to other projects through shared code fragments. This vulnerability is already discussed in section \ref{sec:mostPropagated} in detail including the reason behind its propagation. \textit{CVE-2011-2918} and \textit{CVE-2010-0291} also show a significant propagation, with 358 and 318 occurrences respectively. InterProject vulnerabilities often occur in scenarios where open-source libraries or frameworks are used across different software systems without rigorous security checks. Once a vulnerability is identified in the original project and patched by the vendors, it could have less propagation. But most probably when a vulnerability is found, long before that time this project (whole or maybe part of) have been already reused by several developers. And they would be unaware of the vulnerability announcements and updates, which end up with their vulnerable project. Therefore, these vulnerabilities can easily propagate to any other project that incorporates the same vulnerable code, leading to a chain reaction of security risks.

\figurename~\ref{interprojectSeverity} shows the distribution of InterProject propagated vulnerabilities based on their severity levels, categorized by CVSS scores ranging from 1 to 10. The highest concentration of vulnerabilities falls within the 4.1-5.0 severity range, with 2,359 occurrences, indicating a significant portion of medium severity issues. Following this, 1,594 vulnerabilities are in the 2.1-3.0 range, showing a notable number of low-to-medium severity issues. Severe vulnerabilities (7.1-8.0 range) are also prominent, with 1,151 occurrences, signaling critical risks across different projects. The least frequent are the most critical vulnerabilities (8-10), with only 89 occurrences, suggesting that while they are rare and less, but their impact is potentially devastating and very harmful for your projects. The distribution emphasizes a broad spectrum of risks, with a substantial focus on medium severity vulnerabilities.

\begin{figure}
  \centering
  \includegraphics[scale=.27]{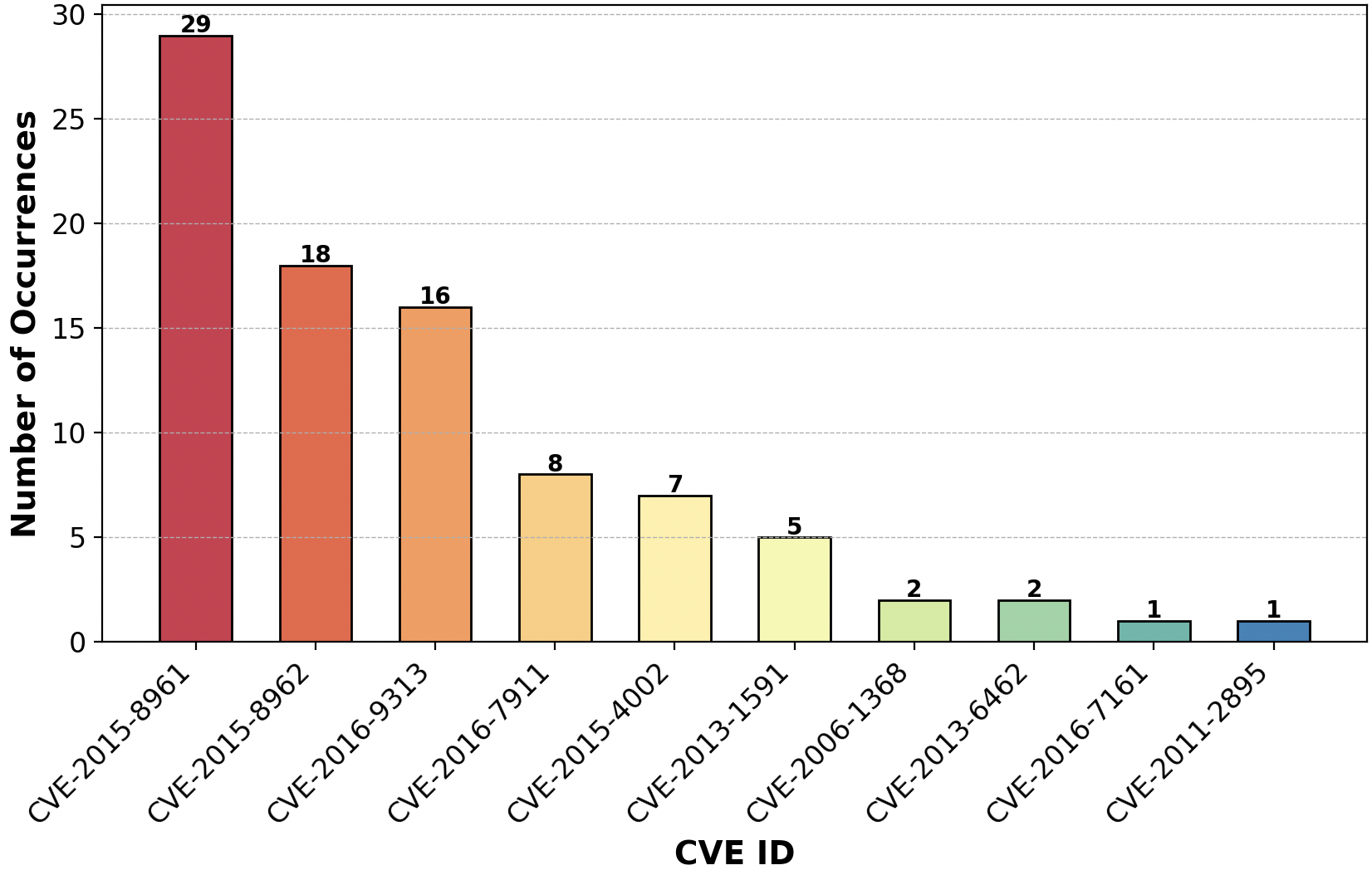}
  \caption{InterProject critical (high risk) propagation detected of range (CVSS: 8-10)}
  \label{interProjectHR}
\end{figure}

\figurename~\ref{interProjectHR} provides a detailed breakdown of critical (high risk) vulnerabilities with CVSS scores between 8 to 10, that have propagated across multiple projects. The most frequently occurring vulnerabilities are \textit{CVE-2015-8961 (29 occurrences)} and \textit{CVE-2015-8962 (18 occurrences)}, both of which are severe and likely to cause significant damage if exploited. \textit{CVE-2015-8961} vulnerability is already explained in section \ref{sec:critical} in detail. \textit{CVE-2016-9313} also stands out with 16 occurrences, indicating its high risk across different projects. The vulnerabilities listed toward the right, such as \textit{CVE-2013-6462}, \textit{CVE-2016-7161} and \textit{CVE-2011-2895}, though less frequent, still represent high-risk issues that require immediate attention of the security maintainers. This figure emphasizes the need for strict security measures to prevent the spread of such critical vulnerabilities across different projects, given their potential to cause a damage.

The detection of such InterProject vulnerabilities highlights the importance of maintaining secure coding practices, regular updates, and thorough vulnerability assessments across all the projects that share code or resources. Moreover, it underscores the need for enhanced collaboration between different development teams and projects to quickly address and mitigate these vulnerabilities before they can cause a widespread damage.

\subsection{Library Used Propagated Vulnerabilities}
At the library-level granularity, we explored another form of vulnerability propagation in our results. In this propagation, various software applications directly or indirectly utilize some open-source third party components or libraries to accelerate their development process. This practice of utilization is quite common, as third-party libraries are designed to be reused. However, it's also common for projects to use outdated or unpatched libraries, which can introduce vulnerabilities or flaws in the system. Although, third-party libraries are an integral part of modern software development and they provide pre-built functionalities to the developers. However, to use these libraries are not out of risks. When the developers incorporate third-party libraries into their software, they also inherit any vulnerabilities those libraries may contain. If a vulnerability exists in a widely-used library, it can propagate across all the projects that use it. Attackers can target a widely-used library to compromise multiple systems. These propagation of vulnerabilities are particularly concerning in open-source libraries, where anyone can contribute. Our tool has detected several vulnerable files (CVEs) in some open source libraries that are often used by the developers. From the results two libraries are shown in Table 1, along with some selected vulnerabilities detected in them.

\begin{table*}
  \centering
  \caption{Discovered vulnerabilities (selected) in different open source libraries}
  \begin{adjustbox}{}
    \begin{tabular}{ccccc}
      \toprule
      Library Name & CVE-ID & CVSS Score & CWE-ID & Vulnerability Type \\
      \midrule
      & CVE-2015-8962 &  9.3  & 415  &  Double Free        \\
      & CVE-2016-7911 &  9.3  & 416  &  Referencing Memory after Freed   \\
      & CVE-2015-4002 &  9  & 119  &  Improper Restriction of Operations        \\
      & CVE-2015-1465 &  7.8  & 17  &  Deprecated Code        \\
      Libos-Nuse
      & CVE-2018-17182 &  7.2  & 416  &  Referencing Memory after Freed         \\
      & CVE-2016-10044 &  7.2  & 264  &  Permissions and Access Controls        \\
      & CVE-2016-10200 &  6.9  & 416  &   Referencing Memory after Freed        \\
      & CVE-2017-15265 &  6.9  & 416  &   Referencing Memory after Freed        \\
      \hline
      & CVE-2014-9659 &   7.5   & 119  & Improper Restriction of Operations   \\
      & CVE-2014-9660 &   7.5   & 476  & NULL Pointer Dereference   \\
      & CVE-2014-9662 &   7.5  & 119  & Improper Restriction of Operation \\
      Libjsqlite
      & CVE-2014-9766 &   7.5  & 189  &  Numeric Errors   \\
      & CVE-2013-6425 &   5    & 189  & Numeric Errors     \\
      & CVE-2014-9675 &   5   & 264  & Permissions and Access Controls   \\
      & CVE-2014-9657 &  7.5  & 125  & Out-of-bounds Read   \\
      \bottomrule
    \end{tabular}
  \end{adjustbox}
\end{table*}

\textbf{LibOS-Nuse} \footnote{https://github.com/libos-nuse/net-next-nuse} :
This library exposes an entry point as API, which is lib\_init(), in order to connect user-space applications to their version of kernel network stack. Once the LibOS is initialized via API, user-space applications can use the system calls defined in LibOS by replacing from the original socket-related symbols to the LibOS specific. We detected more than 15 vulnerabilities of different types in this open source library, few vulnerabilities are mentioned in Table 1, along with the CVSS severity level, CWE id, and the types of vulnerability. Another library \textbf{Libjsqlite} \footnote{https://github.com/geopaparazzi/libjsqlite-spatialite-android} : It is a powerful library that extends the functionality of SQLite with spatial data support for Android applications. It provides a convenient way to store, manage, and query spatial data within your Android apps, enabling you to build location-based features and applications. It allows you to store and manipulate various spatial data types. It can also perform complex spatial queries using SQLite syntax, such as finding nearest neighbors or calculating distances. It works seamlessly with the SQLite database engine, leveraging its efficiency, specifically for Android development. To integrate this library to your project, you need to add it as a dependency to your Android project, create a database, and define spatial tables. Then, you can use SQL queries and spatial functions to interact with your data. Few of the vulnerabilities (CVEs) are shown in the lower part of Table 1, along with the CVSS severity level, CWE id, and the types of vulnerability.

\section{Implications from Our Findings}

\subsection{Summary of Our Findings}
In this section, we summaries our findings including the vulnerability propagation patterns that are detected by our tool SQVDT \cite{akram2021sqvdt}. The discovered vulnerabilities support our hypothesis that how the detected vulnerabilities were propagated from one system to another system. Our analysis revealed distinct propagation patterns for IntraProject and InterProject vulnerabilities. In our analysis, we discovered several InterProject and IntraProject propagated vulnerabilities, which are caused by the sharing of vulnerable code between software projects. The root cause of these propagation often lies in the reuse of insecure software, frameworks, or modules across multiple projects without adequate security audits. Mostly these vulnerabilities propagate either from code clones (copy/paste code), or code reuse in the software. This cloned code could be vulnerable, and the developers use the code without evaluating it. Another common factor of vulnerability propagation is using AI tools to generate the code and use it in your own system without evaluating it. As these tools are trained on some open source repositories like GitHub, on which a lot of software are vulnerable as we discussed in section \ref{sec:Results} with examples. Therefore, these tools also help the vulnerabilities to propagate. For instance in the case of \textit{Library-used} vulnerabilities, the developers frequently integrate open-source components into their projects to accelerate the development process. If these components contain vulnerabilities, any project using them automatically inherits these security flaws. Additionally, the rapid adoption of these components often outpaces the ability of developers to thoroughly check the code for security issues.

\subsection{Vulnerability Management Perspective}
Our findings reveal critical implications for vulnerability management, especially concerning to InterProject and IntraProject, posing significant risks the the systems. This is the same also for vulnerabilities introduced through third-party libraries or AI-generated code, which also unknowingly spread vulnerable code across multiple systems. Proactive monitoring and timely patching are essential to prevent such vulnerabilities from being exploited. Organizations must enforce rigorous security policies to ensure that vulnerabilities are not introduced into their systems. The nature of vulnerability propagation requires a refined approach to risk prioritization. Different code repository handlers (e.g., GitHub) should manage their repositories smartly, so that if the vulnerability detected in one version, all of the previous versions should also be updated automatically.  In cases where InterProject vulnerabilities are identified, collaboration and information sharing with other teams or external organizations is essential. This collective efforts can lead to faster identification and patching of vulnerabilities as well as reducing the overall risks of exploitation.

The lack of coordinated updates and patches across third party libraries, further exacerbates the problem. When a vulnerability is discovered in a shared component, the patching process must be synchronized across all affected projects. This asynchronous response is a critical factor that can minimize the propagation of vulnerabilities from one project to another. An auto patching system could be adopted by organizations to implement comprehensive vulnerability management strategies that include timely updates and coordinated responses across all interrelated projects. The findings underscore the need for a holistic approach to vulnerability management that considers the broader context of software ecosystems. By integrating strategies that address InterProject, IntraProject, third-party library, and AI-generated code vulnerabilities \cite{de2024devaic}, organizations can better safeguard their systems and reduce the overall impact of security threats.

\subsection{Code Clones (Copied Code) Perspective}

During evaluation of vulnerability detection in open source software, we explored that a lot of software are having cloned code from other software, which are exposed as vulnerabilities. We investigate almost 150K wide range of projects (downloaded from GitHub) and analyzed the root cause problem of expanding these vulnerabilities and draw an insight view that these vulnerabilities should be patched as early as possible. Some old vulnerabilities are always tend to remain unpatched in several software systems that can be a copied code from other sources. Mostly these vulnerabilities live for a long time until they were not detected. The cloned vulnerabilities can easily be triggered by some known exploits and can cause a severe damage to the system. Patching these vulnerabilities is very challenging when multiple vendors are involved in this copy/paste chain. Furthermore, it's very hard to manually patch every cloned venerability at its code level. According to result analysis, we can classified these vulnerabilities into two categories, InterProject, IntraProject.

Code clones significantly influence vulnerability propagation across software projects. Cloned code can exist across multiple projects (InterProject clones) or within different components of the same project (IntraProject clones). When vulnerabilities are introduced in a piece of code, they are often duplicated wherever the code is cloned, leading to widespread security risks. InterProject propagation can be particularly problematic if the vulnerabilities are not recognized and patched in all instances of the cloned code. Within a same project, developers often clone code to reuse functionalities into the extended version. Although, this can speed up development but it also means that any existing vulnerabilities in the original code are replicated across the project. Meanwhile, AI-generated code can further complicate the issue of code clones. As the AI tools often generate code based on the existing patterns or codebases, they may inadvertently introduce or replicate vulnerabilities found in their training data. This can lead to the propagation of insecure coding practices across both IntraProject and InterProject clones. Automated tools for detecting code clones and their associated vulnerabilities can help to identify and mitigate risks. Regular code reviews and consistent security testing are essential to manage the vulnerability propagation through cloned code. By understanding and addressing the implications of InterProject and IntraProject code cloning, alongside the challenges posed by AI-generated code, organizations can better protect their software systems from vulnerability propagation. A study related to vulnerability detection in AI-generated code has been proposed by Domenico et al. \cite{de2024devaic} in very detail.

\subsection{Third-Party Library used Perspective}
A lot of vulnerabilities propagate just because the use of open source libraries, and there could be several causes of this. Modern software often relies on a complex network of dependencies. A vulnerability in one library can spread to others that depend on it, creating a chain reaction. For instance, if a foundational library like OpenSSL contains a flaw, it can affect all software that relies on it directly or indirectly. Even when vulnerabilities are discovered in a library and patches are released, not all projects related to these libraries adopt these patches immediately. This delay can leave systems exposed to known vulnerabilities, especially if the library is deeply integrated into the software. Sometimes, developers may not always be aware of the vulnerabilities in the libraries they use. This lack of awareness can lead to the unintentional inclusion of vulnerable code in their software. Several popular libraries are often used across numerous projects and industries. A vulnerability in such a library can have far-reaching consequences, affecting critical systems across the globe.

Our finding on the propagation of vulnerabilities through third-party libraries highlight critical implications for organizations. Organizations need to integrate some vulnerability scanning tools into their development pipelines to detect potential issues early. This approach can prevent vulnerabilities from being introduced into production environments. Vulnerabilities in the third-party libraries often receive public attention quickly, and the time span between a vulnerability's disclosure and its exploitation can be narrow. Organizations that delay in applying security patches may find themselves exposed to the significant risks. By concentrating on the most critical vulnerabilities, the organizations can make more efficient use of their security resources. Finally, collaboration with vendors of third-party libraries is quite important. In cases where vulnerabilities are identified in the libraries, we should be working closely with the vendor to ensure the patching and updating.

\subsection{Systemic Issues behind Propagation}
Systemic issues refer to the underlying, pervasive problems within a software ecosystem that allow certain vulnerabilities to persist and propagate across multiple systems, projects, or environments. These issues are often deep-rooted, reflecting broader challenges in how software is developed, maintained, and secured. The presence of older vulnerabilities in the top 20 most propagated CVEs as shown in \figurename~\ref{topCVE}, \ref{linux20}, \ref{android20}, \ref{interProject20}, is a strong indicator of systemic issues. Vulnerability propagation at this much large scale could be because of the following factors, which are also be the root causes of this propagation. Many open-source projects rely on third-party libraries, frameworks, or components that might contain vulnerabilities. When a project continues to use outdated versions of these dependencies, known vulnerabilities (such as those from 2010 or 2011) can persist, even though patches or updates are already available. This suggests a systemic issue where projects either do not prioritize updates or maybe lack the resources to do so. The presence of long-known vulnerabilities like CVE-2010-0291 and CVE-2011-2918 in the top 20 list (as shown in \figurename~\ref{topCVE}, \ref{linux20}, and \ref{interProject20}), indicates that many systems have not been applied the available patches. This could be due to lack of awareness, compatibility concerns, or resource constraints. When a particular software component is widely adopted across multiple projects, any vulnerability within it has the potential to propagate extensively. The systemic issue here lies in the widespread use of shared, yet vulnerable, components without adequate governance or updating mechanisms. Additionally, the persistence of old vulnerabilities also points to a lack of robust security practices within the development life-cycle. This might include inadequate vulnerability scanning, poor code review processes, or insufficient security training for developers. If developers and maintainers are not equipped to identify and address vulnerabilities, these issues can persist across multiple versions of projects. Many organizations continue to use legacy systems or software that are no longer actively maintained as discussed in section \ref{sec:Results}. These systems are often more vulnerable because they do not receive regular updates or patches. As a result, vulnerabilities remain unaddressed, creating security risks across the ecosystem. In some cases, vulnerabilities are not disclosed or identified until long after they have been introduced. This delay in discovery means that by the time a vulnerability is known, it may have already propagated widely across various projects.

\section{Related Work} \label{sec:Related}
The study of software vulnerabilities, particularly their propagation from one system to another, has been an area of growing research interest as interdependence of modern software continue to increase. This research has expanded into various domains, from understanding the spread of vulnerabilities within a single project (intra-project) to the propagation between different projects (inter-project), the risks associated with third-party libraries, and even vulnerabilities in AI-generated code. We found several studies related to the vulnerability detection in the code or in the third-party components and removing code clones that propagate the software bugs to other systems. But we didn't find any study that talks about the CVE based vulnerability propagation from one system to another system.

A lot of researchers have been working on the prediction of vulnerabilities by using modeling techniques, in which they collected the data of previously published CVEs and on the basis of this dataset they predict future occurring of the vulnerability \cite{khoury2021analysis}. The growth model of software reliability relate the number of defects during testing time. In the most recent work, it has been focused on detecting vulnerabilities in OSS (open source software), meanwhile proposing the prediction models for future occurring vulnerabilities, which do not concern with source code but only to the CVE data \cite{jabeen2019integrated, kim2013should, nappa2015attack, kim2018software}. Detecting code clones in big code bases have been very useful to overcome security flaws \cite{juergens2009clonedetective, sajnani2015parallel, akram2018DCCD}. Prior research has also shown that there are many vulnerable applications in android markets \cite{Akram2019DroidMD, chen2014achieving, zhou2012detecting} having malware in their apk files, These applications are being used by different users without knowing that if it is vulnerable. Chengwei et al. \cite{liu2022demystifying} proposed a graph based dependency resolution, which resolves the inner dependency relations as trees (dependency trees), and investigates the threats from vulnerabilities in dependency trees. The security threats were introduced by vulnerabilities in the third-party JavaScript libraries within the NPM ecosystem.

There are different tools available in the market that could detect vulnerabilities in the code but have some limitation. For example, ReDeBug \cite{jang2012redebug} takes a set of lines of code as its processing unit and slides a window of \textit{n = 4} through the source code and applies hash functions to each window. The clones detected by bloom filter, which save the hashes of every window. ReDeBug detects some of type-3 clones, which are the indication of more false rate. VFDETECT \cite{liu2017vfdetect} proposed an approach based on innovative fingerprint model to detect vulnerabilities in the source code. VFDETECT could maintain better performance and acquire higher robustness under multiple code modification methods such as variable renaming, code sequence changing and redundancy inserting, which is difficult to achieve in existing research work. Our results in real-world datasets detection demonstrate that it is of practical values. DECKARD \cite{jiang2007deckard} extracts all the characteristic vectors from ASTs (for every file). This approach is a bit slow and require extensive time sub-graph isomorphism. Meanwhile DECKARD does not guarantee the scalability. Moreover, DECKARD has 90\% false positive rate, which is not acceptable for the detection of exact vulnerable files. VulPecker \cite{li2016vulpecker} is an automatically checking system and it characterizes all the vulnerabilities with different set of features, after that it selects one code optimal similarity algorithms for vulnerability detection. It has detected almost 40 vulnerabilities, which were not registered in the NVD. But still its not good for massive vulnerability detection in open source software, because it is slow.

Effective security strategies are required, as any vulnerability in the system could generate severe financial implications in local systems as well as in industrial systems \cite{chen2020survey, lin2023coordination, aslan2023comprehensive, ma2020impact, alfadel2023empirical}. Darshana et al. \cite{upadhyay2020scada} proposed a system SCADA to identify potential breaches or aspects in the system where a breach may occur. Authors describe various types of vulnerabilities by taking real incidents that were reported in the standard vulnerability databases. Iannone et al. \cite{iannone2021toward} introduce SIEGE, a genetic algorithms based novel automatic exploit generation approach it generates test cases that execute methods within a library known to contain a vulnerability. Additionally, they guide security researchers in understanding how the vulnerability could be exploited in practice. Xingbin et al. \cite{jiang2020experimental} examine the attack surfaces of a networked embedded system, composed of devices used in the IIoT field. Authors set up a concrete attack to gain control of the system actuators and perform some hazardous operations.
Zhen et al. \cite{li2021vuldeelocator} present VulDeeLocator a vulnerability detector that can achieve a high detection capability along with high precision. Yueming et al. \cite{wu2022vulcnn} propose a novel idea that can efficiently convert the source code of a function into an image while preserving the details. They implement VulCNN and evaluate it on a dataset of 13,687 vulnerable functions, the results report that VulCNN can achieve better accuracy than state-of-the-art tools.

Code clones also becoming the cause of vulnerability propagation from one system to another system. A lot techniques are available, which detects code clones in the projects and guide users to remove those clones to maintain code quality. Debarshi et al. \cite{chatterji2011measuring} investigates the usefulness of code clone information and perform a bug localization task. The goal was to understand how developers used clone information to perform this task. The results of the study showed that participants who first identified a defect then used it to look for clones of the defect were more effective than participants who used the clone information before finding any defects. This study also show a relationship between the efficacy of the clone information and effectiveness in finding defects. Matthew et al. \cite{beard2011extending} examine the application of code clone location techniques in the context of bug localization. The authors determine the advantages of extending existing code clone location techniques through the inclusion of models in the analysis process. They also examine a technique for extending the use of bug logging repositories. In some studies, the security vulnerability propagation analysis is performed to determine the propagation paths with the highest probability and the largest estimated risk value. For example, Feng et al. \cite{feng2014security} claims that security vulnerability could also propagate through a causal chains of risk factors via multiple paths, They proposed a model (SRAM) to simultaneously define the risk factors and their causal relationships based on the knowledge from observed cases and domain experts. Then, the analysis is performed to determine the propagation paths with the highest probability SRAM enables organizations to establish security risk management plans.

If we talk about AI code generators, many researchers are working on the training of AI models to developing tools which evaluate the AI-generated code \cite{yeticstiren2023evaluating, hamer2024just, al2023chatgpt, negri2024systematic}. For example, DeVAIC \cite{cotroneo2024devaic} is a methodological approach based tool to evaluate the AI-generated Python code, it involves gathering the vulnerable samples, extracting their patterns, and evaluating the code. Kapitsaki et al. \cite{kapitsaki2024generative} used five tools: OpenAI ChatGPT, Google Gemini, GitHub Copilot, TabNine and Amazon CodeWhisperer, and provide an overview of the tools when examining code generation. Their work introduces a conceptual model that can help to improve the software reuse factor. Tomas et al. \cite{mateo2024enhancing} demonstrate how self-generative vulnerabilities manifest in programming, through a case study. The proposed work illustrates a practical example of software vulnerability existing in the AI generated code using ChatGPT. Additionally, they conduct an experiment that generate SQL injection attack, by highlighting the hacker’s actions to exploit that generated vulnerability. They recommended an effective prevention strategy that includes the training programs and error analysis in software development.

\section{Threats to Validity}\label{sec:Validity}

In this section, we discuss the main threats to the validity to our findings in this research work. We categorize the threats into construct validity, internal validity, external validity, and statistical conclusion by considering the potential limitations and the measures taken to mitigate them.

\textbf{Construction Validity:} One potential threat arises from the interpretation of propagation when vulnerabilities involve patches affecting a large number of files. For example, vulnerabilities such as CVE-2013-7421 required modifications across many files, including numerous insertions and deletions. In such cases, if only a subset of the modified files is reused in another project, it may not necessarily reproduce the original vulnerability context. To mitigate this threat, our analysis focuses on vulnerable code fragments associated with specific CVEs, rather than treating every reused file involved in a patch as a propagation instance. Additionally, the SQVDT framework relies on vulnerability signatures derived from known CVEs and their patches, which helps ensure that detected matches correspond to actual vulnerable code patterns rather than generic code similarities.

\textbf{Internal Validity:} One possible threat is that the detected similarities between code fragments may arise from common programming practices, shared libraries, or independent implementation of similar functionality, rather than direct reuse of vulnerable code. To reduce this risk, SQVDT focuses specifically on vulnerable code fragments linked to documented CVEs, using patch-based vulnerability signatures rather than generic similarity measures. This approach reduces the likelihood of incorrectly identifying unrelated code fragments as propagated vulnerabilities. Furthermore, the tool itself has been previously evaluated in our earlier work, where its detection capability and accuracy metrics were validated, demonstrating a very low false-positive rate.

\textbf{External Validity:} One potential limitation of this study is related to data completeness. The dataset of open-source projects was randomly collected from GitHub and contains approximately 150,000 projects, which represents only a small portion of the large number of repositories available on the platform. Similarly, the vulnerability dataset used in this study includes 1,870 CVEs, which may not represent the entire spectrum of software vulnerabilities. Another factor affecting generalizability is the rapid evolution of software systems and vulnerabilities. Software projects continuously evolve through patches, updates, and refactoring, meaning that some vulnerabilities identified in earlier versions may already be fixed in later releases. As a result, the findings presented in this study reflect the state of the analyzed systems during the time of data collection.

Additionally, the analysis of AI-generated code vulnerabilities in this study is limited to a small exploratory experiment. The objective of this analysis was to illustrate the possibility that vulnerable patterns may appear in automatically generated code. However, a comprehensive evaluation of vulnerabilities in AI-generated code would require large-scale datasets, systematic experimentation across multiple AI models, and extensive vulnerability validation. Therefore, the results presented in this section should be interpreted as indicative rather than conclusive, and further research is needed to confirm these observations.

\textbf{Statistical Conclusion:} Since this research primarily focuses on empirical observations of vulnerability propagation patterns, rather than complex statistical hypothesis testing, the risk of incorrect statistical inference is relatively limited. Nevertheless, potential bias may arise from dataset size and sampling limitations. To mitigate this issue, we analyzed a diverse set of open-source systems and vulnerabilities obtained from publicly available vulnerability databases. Moreover, the analysis emphasizes observable propagation patterns rather than statistical generalization, which helps reduce the risk of misleading conclusions.

\section{Conclusion}\label{sec:conclusion}
The growing complexity of software systems and the increasing reliance on open-source software, third-party components have significantly amplified the challenges associated with software security. Our research highlights the persistent and evolving nature of software vulnerabilities propagation and emphasize the critical need for effective vulnerability management strategies. By using SQVDT tool, we have identified the propagation of vulnerabilities, both IntraProject and InterProject, as a significant threat to software security. We have also listed several vulnerabilities that were detected in different open source projects and third-party libraries along with their CVSS severity level, CWE id, and the type. We found several vulnerabilities that were propagated only because reusing the code through cloning techniques and the integration of third-party libraries into different projects.  There is a need for further research into the mechanisms of vulnerability propagation, with an emphasis on developing automated solutions to identify and mitigate these risks. Additionally, A big need is to analyze some AI-generated code fragments, that are causing the vulnerability propagation. Future work should focus on enhancing vulnerability detection tools, particularly for AI-generated code, and developing a robust strategy for managing vulnerabilities in open-source software.

\bibliography{wileyNJD-Chicago}

@article{akram2021sqvdt,
  title={SQVDT: A scalable quantitative vulnerability detection technique for source code security assessment},
  author={Akram, Junaid and Luo, Ping},
  journal={Software: Practice and Experience},
  volume={51},
  number={2},
  pages={294--318},
  year={2021},
  publisher={Wiley Online Library}
}

@inproceedings{akram2018droidcc,
  title={DroidCC: A Scalable Clone Detection Approach for Android Applications to Detect Similarity at Source Code Level},
  author={Akram, Junaid and Shi, Zhendong and Mumtaz, Majid and Luo, Ping},
  booktitle={2018 IEEE 42nd Annual Computer Software and Applications Conference (COMPSAC)},
  pages={100--105},
  year={2018},
  organization={IEEE}
}

@inproceedings{akram2018DCCD,
  title={DCCD: An Efficient and Scalable Distributed Code Clone Detection Technique for Big Code},
  author={Akram, Junaid and Shi, Zhendong and Mumtaz, Majid and Luo, Ping},
  booktitle={PROCEEDINGS SEKE 2018, The 30 th International Conference on Software Engineering and Knowledge Engineering},
  pages={354--359},
  year={2018},
  organization={KSI Research Inc}
}

@article{akram2019DroidMD,
  title={DroidMD: An Efficient and Scalable Android Malware Detection Approach at Source Code Level},
  author={Akram, Junaid and Mumtaz, Majid and Gul, Jabeen and Luo, Ping},
  journal={International Journal of Information and Computer Security},
  doi = {10.1504/IJICS.2019.10020453},
  volume={11},
  number={1},
  year={2019},
  publisher={IJICS}
}

@inproceedings{akram2019VCIPR,
  title={VCIPR : Vulnerable Code is Identifiable When a Patch is Released (Hacker's Perspective)},
  author={Akram, Junaid and Liang, Qi and Ping, Luo},
  doi = {10.1109/ICST.2019.00049},
  booktitle={2019 12th IEEE Conference on Software Testing, Validation and Verification (ICST)},
  pages={402--413},
  year={2019},
  organization={IEEE}
}

@article{Akram2019how,
  title={How to build a vulnerability benchmark to overcome cyber security attacks},
  author={Akram, Junaid and Ping, Luo},
  journal={IET Information Security},
  volume={14},
  number={1},
  pages={60--71},
  year={2019},
  publisher={IET}
}

@article{Akram2019IBFET,
  title={IBFET: Index-based features extraction technique for scalable code clone detection at file level granularity},
  author={Akram, Junaid and Mumtaz, Majid and Luo, Ping},
  journal={Software: Practice and Experience},
  volume={50},
  number={1},
  pages={22--46},
  year={2020},
  publisher={Wiley Online Library}
}

@inproceedings{mumtaz2019rsa,
  title={An RSA based authentication system for smart IoT environment},
  author={Mumtaz, Majid and Akram, Junaid and Ping, Luo},
  booktitle={2019 IEEE 21st International Conference on High Performance Computing and Communications; IEEE 17th International Conference on Smart City; IEEE 5th International Conference on Data Science and Systems (HPCC/SmartCity/DSS)},
  pages={758--765},
  year={2019},
  organization={IEEE}
}

@inproceedings{jabeen2019integrated,
  title={An Integrated Software Vulnerability Discovery Model based on Artificial Neural Network.},
  author={Jabeen, Gul and Ping, Luo and Akram, Junaid and Shah, Akber Aman},
  booktitle={SEKE},
  pages={349--458},
  year={2019},
  organization={SEKE: Software Engineering and Knowledge Engineering}
}

@article{avery2018formally,
  title={Formally modeling deceptive patches using a game-based approach},
  author={Avery, Jeffrey and Wallrabenstein, John Ross},
  journal={Computers \& Security},
  volume={75},
  pages={182--190},
  year={2018},
  publisher={Elsevier}
}

@article{alhazmi2007measuring,
  title={Measuring, analyzing and predicting security vulnerabilities in software systems},
  author={Alhazmi, Omar H and Malaiya, Yashwant K and Ray, Indrajit},
  journal={Computers \& Security},
  volume={26},
  number={3},
  pages={219--228},
  year={2007},
  publisher={Elsevier}
}

@inproceedings{gorbenko2017experience,
  title={Experience Report: Study of Vulnerabilities of Enterprise Operating Systems},
  author={Gorbenko, Anatoliy and Romanovsky, Alexander and Tarasyuk, Olga and Biloborodov, Oleksandr},
  booktitle={Software Reliability Engineering (ISSRE), 2017 IEEE 28th International Symposium on},
  pages={205--215},
  year={2017},
  organization={IEEE}
}

@inproceedings{guo2016research,
  title={Research on Detecting Windows Vulnerabilities Based on Security Patch Comparison},
  author={Guo, Hui and Wang, Yong-Yi and Pan, Zu-Lie and Liu, Shi-Wei},
  booktitle={Instrumentation \& Measurement, Computer, Communication and Control (IMCCC), 2016 Sixth International Conference on},
  pages={366--369},
  year={2016},
  organization={IEEE}
}

@inproceedings{jang2012redebug,
  title={ReDeBug: finding unpatched code clones in entire os distributions},
  author={Jang, Jiyong and Agrawal, Abeer and Brumley, David},
  booktitle={Security and Privacy (SP), 2012 IEEE Symposium on},
  pages={48--62},
  year={2012},
  organization={IEEE}
}

@inproceedings{kim2017vuddy,
  title={VUDDY: A Scalable Approach for Vulnerable Code Clone Discovery},
  author={Kim, Seulbae and Woo, Seunghoon and Lee, Heejo and Oh, Hakjoo},
  booktitle={Security and Privacy (SP), 2017 IEEE Symposium on},
  pages={595--614},
  year={2017},
  organization={IEEE}
}

@inproceedings{brumley2008automatic,
  title={Automatic patch-based exploit generation is possible: Techniques and implications},
  author={Brumley, David and Poosankam, Pongsin and Song, Dawn and Zheng, Jiang},
  booktitle={Security and Privacy, 2008. SP 2008. IEEE Symposium on},
  pages={143--157},
  year={2008},
  organization={IEEE}
}

@article{bojanova2023heartbleed,
  title={Heartbleed revisited: Is it just a buffer over-read?},
  author={Bojanova, Irena and Galhardo, Carlos Eduardo C},
  journal={IT Professional},
  volume={25},
  number={2},
  pages={83--89},
  year={2023},
  publisher={IEEE}
}

@misc{techtarget,
  author = {Sean Michael Kerner},
  title = {35 cybersecurity statistics to lose sleep over in 2024},
  howpublished = "\url{https://www.techtarget.com/whatis/34-Cybersecurity-Statistics-to-Lose-Sleep-Over-in-2020}",
    note = "35 cybersecurity statistics to lose sleep over in 2024",
   year = {Online accessed: 19-August-2024}
}

@article{kim2013should,
  title={Where should we fix this bug? a two-phase recommendation model},
  author={Kim, Dongsun and Tao, Yida and Kim, Sunghun and Zeller, Andreas},
  journal={IEEE transactions on software Engineering},
  volume={39},
  number={11},
  pages={1597--1610},
  year={2013},
  publisher={IEEE}
}

@inproceedings{aggarwal2023study,
  title={A Study of CVSS v4. 0: A CVE Scoring System},
  author={Aggarwal, Manuj},
  booktitle={2023 6th International Conference on Contemporary Computing and Informatics (IC3I)},
  volume={6},
  pages={1180--1186},
  year={2023},
  organization={IEEE}
}

@inproceedings{nappa2015attack,
  title={The attack of the clones: A study of the impact of shared code on vulnerability patching},
  author={Nappa, Antonio and Johnson, Richard and Bilge, Leyla and Caballero, Juan and Dumitras, Tudor},
  booktitle={Security and Privacy (SP), 2015 IEEE Symposium on},
  pages={692--708},
  year={2015},
  organization={IEEE}
}

@article{kim2018software,
  title={Software systems at risk: An empirical study of cloned vulnerabilities in practice},
  author={Kim, Seulbae and Lee, Heejo},
  journal={Computers \& Security},
  year={2018},
  publisher={Elsevier}
}

@inproceedings{liu2017vfdetect,
  title={VFDETECT: A vulnerable code clone detection system based on vulnerability fingerprint},
  author={Liu, Zhen and Wei, Qiang and Cao, Yan},
  booktitle={Information Technology and Mechatronics Engineering Conference (ITOEC), 2017 IEEE 3rd},
  pages={548--553},
  year={2017},
  organization={IEEE}
}

@inproceedings{ng2014android,
  title={Which android app store can be trusted in China?},
  author={Ng, Yi Ying and Zhou, Hucheng and Ji, Zhiyuan and Luo, Huan and Dong, Yuan},
  booktitle={Computer Software and Applications Conference (COMPSAC), 2014 IEEE 38th Annual},
  pages={509--518},
  year={2014},
  organization={IEEE}
}

@inproceedings{juergens2009code,
  title={Do code clones matter?},
  author={Juergens, Elmar and Deissenboeck, Florian and Hummel, Benjamin and Wagner, Stefan},
  booktitle={Software Engineering, 2009. ICSE 2009. IEEE 31st International Conference on},
  pages={485--495},
  year={2009},
  organization={IEEE}
}

@article{sheneamer2016survey,
  title={A Survey of Software Clone Detection Techniques},
  author={Sheneamer, Abdullah and Kalita, Jugal},
  journal={International Journal of Computer Applications},
  pages={0975--8887},
  year={2016}
}

@inproceedings{juergens2009clonedetective,
  title={CloneDetective-A workbench for clone detection research},
  author={Juergens, Elmar and Deissenboeck, Florian and Hummel, Benjamin},
  booktitle={Proceedings of the 31st International Conference on Software Engineering},
  pages={603--606},
  year={2009},
  organization={IEEE Computer Society}
}

@article{sajnani2015parallel,
  title={A parallel and efficient approach to large scale clone detection},
  author={Sajnani, Hitesh and Saini, Vaibhav and Lopes, Cristina},
  journal={Journal of Software: Evolution and Process},
  volume={27},
  number={6},
  pages={402--429},
  year={2015},
  publisher={Wiley Online Library}
}

@inproceedings{chen2014achieving,
  title={Achieving accuracy and scalability simultaneously in detecting application clones on android markets},
  author={Chen, Kai and Liu, Peng and Zhang, Yingjun},
  booktitle={Proceedings of the 36th International Conference on Software Engineering},
  pages={175--186},
  year={2014},
  organization={ACM}
}

@inproceedings{zhou2012detecting,
  title={Detecting repackaged smartphone applications in third-party android marketplaces},
  author={Zhou, Wu and Zhou, Yajin and Jiang, Xuxian and Ning, Peng},
  booktitle={Proceedings of the second ACM conference on Data and Application Security and Privacy},
  pages={317--326},
  year={2012},
  organization={ACM}
}

@inproceedings{jiang2007deckard,
  title={Deckard: Scalable and accurate tree-based detection of code clones},
  author={Jiang, Lingxiao and Misherghi, Ghassan and Su, Zhendong and Glondu, Stephane},
  booktitle={Proceedings of the 29th international conference on Software Engineering},
  pages={96--105},
  year={2007},
  organization={IEEE Computer Society}
}

@inproceedings{li2016vulpecker,
  title={VulPecker: an automated vulnerability detection system based on code similarity analysis},
  author={Li, Zhen and Zou, Deqing and Xu, Shouhuai and Jin, Hai and Qi, Hanchao and Hu, Jie},
  booktitle={Proceedings of the 32nd Annual Conference on Computer Security Applications},
  pages={201--213},
  year={2016},
  organization={ACM}
}

@misc{lawler2017equifax,
  author = {Lawler, Richard},
  title = {Equifax Blames Breach on a Server Flaw It Should Have Patched},
  year = {2017},
  month = sep,
  howpublished = {Engadget},
  note = {Published September 13, 2017. Available: \url{https://www.engadget.com/2017-09-13-equifax-apache-argentina.html}},
  urldate = {2026-08-14}
}

@misc{nvdCVE20181274,
  author = {{National Institute of Standards and Technology}},
  title = {{CVE-2018-1274 Detail}},
  year = {2018},
  howpublished = {National Vulnerability Database},
  note = {Available: \url{https://nvd.nist.gov/vuln/detail/CVE-2018-1274}},
  urldate = {2026-08-14}
}

@article{alqahtani2016tracing,
  title={Tracing known security vulnerabilities in software repositories--A Semantic Web enabled modeling approach},
  author={Alqahtani, Sultan S and Eghan, Ellis E and Rilling, Juergen},
  journal={Science of Computer Programming},
  volume={121},
  pages={153--175},
  year={2016},
  publisher={Elsevier}
}

@inproceedings{khoury2021analysis,
  title={An analysis of the use of CVEs by IoT malware},
  author={Khoury, Rapha{\"e}l and Vignau, Benjamin and Hall{\'e}, Sylvain and Hamou-Lhadj, Abdelwahab and Razgallah, Asma},
  booktitle={Foundations and Practice of Security: 13th International Symposium, FPS 2020, Montreal, QC, Canada, December 1--3, 2020, Revised Selected Papers 13},
  pages={47--62},
  year={2021},
  organization={Springer}
}

@inproceedings{ma2020impact,
  title={Impact analysis of cross-project bugs on software ecosystems},
  author={Ma, Wanwangying and Chen, Lin and Zhang, Xiangyu and Feng, Yang and Xu, Zhaogui and Chen, Zhifei and Zhou, Yuming and Xu, Baowen},
  booktitle={Proceedings of the ACM/IEEE 42nd International Conference on Software Engineering},
  pages={100--111},
  year={2020},
  organization={IEEE}
}

@inproceedings{chatterji2011measuring,
  title={Measuring the efficacy of code clone information in a bug localization task: An empirical study},
  author={Chatterji, Debarshi and Carver, Jeffrey C and Massengil, Beverly and Oslin, Jason and Kraft, Nicholas A},
  booktitle={2011 International Symposium on Empirical Software Engineering and Measurement},
  pages={20--29},
  year={2011},
  organization={IEEE}
}

@inproceedings{beard2011extending,
  title={Extending bug localization using information retrieval and code clone location techniques},
  author={Beard, Matthew},
  booktitle={2011 18th Working Conference on Reverse Engineering},
  pages={425--428},
  year={2011},
  organization={IEEE}
}

@article{feng2014security,
  title={A security risk analysis model for information systems: Causal relationships of risk factors and vulnerability propagation analysis},
  author={Feng, Nan and Wang, Harry Jiannan and Li, Minqiang},
  journal={Information sciences},
  volume={256},
  pages={57--73},
  year={2014},
  publisher={Elsevier}
}

@inproceedings{liu2022demystifying,
  title={Demystifying the vulnerability propagation and its evolution via dependency trees in the npm ecosystem},
  author={Liu, Chengwei and Chen, Sen and Fan, Lingling and Chen, Bihuan and Liu, Yang and Peng, Xin},
  booktitle={Proceedings of the 44th International Conference on Software Engineering},
  pages={672--684},
  year={2022},
  organization={ IEEE/ACM}
}

@inproceedings{iannone2021toward,
  title={Toward automated exploit generation for known vulnerabilities in open-source libraries},
  author={Iannone, Emanuele and Di Nucci, Dario and Sabetta, Antonino and De Lucia, Andrea},
  booktitle={2021 IEEE/ACM 29th International Conference on Program Comprehension (ICPC)},
  pages={396--400},
  year={2021},
  organization={IEEE}
}

@inproceedings{cotroneo2024vulnerabilities,
  title={Vulnerabilities in ai code generators: Exploring targeted data poisoning attacks},
  author={Cotroneo, Domenico and Improta, Cristina and Liguori, Pietro and Natella, Roberto},
  booktitle={Proceedings of the 32nd IEEE/ACM International Conference on Program Comprehension},
  pages={280--292},
  year={2024},
  organization={IEEE/ACM}
}

@article{cotroneo2024devaic,
  title={Devaic: A tool for security assessment of ai-generated code},
  author={Cotroneo, Domenico and De Luca, Roberta and Liguori, Pietro},
  journal={arXiv preprint arXiv:2404.07548},
  year={2024}
}

@article{yeticstiren2023evaluating,
  title={Evaluating the code quality of ai-assisted code generation tools: An empirical study on github copilot, amazon codewhisperer, and chatgpt},
  author={Yeti{\c{s}}tiren, Burak and {\"O}zsoy, I{\c{s}}{\i}k and Ayerdem, Miray and T{\"u}z{\"u}n, Eray},
  journal={arXiv preprint arXiv:2304.10778},
  year={2023}
}

@inproceedings{kapitsaki2024generative,
  title={Generative AI for Code Generation: Software Reuse Implications},
  author={Kapitsaki, Georgia M},
  booktitle={International Conference on Software and Software Reuse},
  pages={37--47},
  year={2024},
  organization={Springer}
}

@inproceedings{hamer2024just,
  title={Just another copy and paste? Comparing the security vulnerabilities of ChatGPT generated code and StackOverflow answers},
  author={Hamer, Sivana and d’Amorim, Marcelo and Williams, Laurie},
  booktitle={2024 IEEE Security and Privacy Workshops (SPW)},
  pages={87--94},
  year={2024},
  organization={IEEE}
}

@article{chen2020survey,
  title={A survey on ethereum systems security: Vulnerabilities, attacks, and defenses},
  author={Chen, Huashan and Pendleton, Marcus and Njilla, Laurent and Xu, Shouhuai},
  journal={ACM Computing Surveys (CSUR)},
  volume={53},
  number={3},
  pages={1--43},
  year={2020},
  publisher={ACM New York, NY, USA}
}

@article{upadhyay2020scada,
  title={SCADA (Supervisory Control and Data Acquisition) systems: Vulnerability assessment and security recommendations},
  author={Upadhyay, Darshana and Sampalli, Srinivas},
  journal={Computers \& Security},
  volume={89},
  pages={101666},
  year={2020},
  publisher={Elsevier}
}

@article{jiang2020experimental,
  title={An experimental analysis of security vulnerabilities in industrial IoT devices},
  author={Jiang, Xingbin and Lora, Michele and Chattopadhyay, Sudipta},
  journal={ACM Transactions on Internet Technology (TOIT)},
  volume={20},
  number={2},
  pages={1--24},
  year={2020},
  publisher={ACM New York, NY, USA}
}

@article{li2021vuldeelocator,
  title={Vuldeelocator: a deep learning-based fine-grained vulnerability detector},
  author={Li, Zhen and Zou, Deqing and Xu, Shouhuai and Chen, Zhaoxuan and Zhu, Yawei and Jin, Hai},
  journal={IEEE Transactions on Dependable and Secure Computing},
  volume={19},
  number={4},
  pages={2821--2837},
  year={2021},
  publisher={IEEE}
}

@inproceedings{wu2022vulcnn,
  title={Vulcnn: An image-inspired scalable vulnerability detection system},
  author={Wu, Yueming and Zou, Deqing and Dou, Shihan and Yang, Wei and Xu, Duo and Jin, Hai},
  booktitle={Proceedings of the 44th International Conference on Software Engineering},
  pages={2365--2376},
  year={2022},
  organization = {IEEE/ACM}
}

@article{lin2023coordination,
  title={On the coordination of vulnerability fixes: An empirical study of practices from 13 CVE numbering authorities},
  author={Lin, Jiahuei and Adams, Bram and Hassan, Ahmed E},
  journal={Empirical Software Engineering},
  volume={28},
  number={6},
  pages={151},
  year={2023},
  publisher={Springer}
}

@article{aslan2023comprehensive,
  title={A comprehensive review of cyber security vulnerabilities, threats, attacks, and solutions},
  author={Aslan, {\"O}mer and Aktu{\u{g}}, Semih Serkant and Ozkan-Okay, Merve and Yilmaz, Abdullah Asim and Akin, Erdal},
  journal={Electronics},
  volume={12},
  number={6},
  pages={1333},
  year={2023},
  publisher={MDPI}
}

@article{alfadel2023empirical,
  title={Empirical analysis of security vulnerabilities in python packages},
  author={Alfadel, Mahmoud and Costa, Diego Elias and Shihab, Emad},
  journal={Empirical Software Engineering},
  volume={28},
  number={3},
  pages={59},
  year={2023},
  publisher={Springer}
}

@article{mateo2024enhancing,
  title={Enhancing Security in Industrial Application Development: Case Study on Self-Generating Artificial Intelligence Tools},
  author={Mateo Sanguino, Tom{\'a}s de J},
  journal={Applied Sciences},
  volume={14},
  number={9},
  pages={3780},
  year={2024},
  publisher={MDPI}
}

@article{negri2024systematic,
  title={A systematic literature review on the impact of AI models on the security of code generation},
  author={Negri-Ribalta, Claudia and Geraud-Stewart, R{\'e}mi and Sergeeva, Anastasia and Lenzini, Gabriele},
  journal={Frontiers in Big Data},
  volume={7},
  pages={1386720},
  year={2024},
  publisher={Frontiers Media SA}
}

@mastersthesis{syvertsen2011selection,
  title={Selection and use of third-party Software Components: Study of a IT consultancy firm},
  author={Syvertsen, Martin M},
  year={2011},
  school={Institutt for datateknikk og informasjonsvitenskap},
  address = {Norwegian University of Science and Technology}
}

@article{al2023chatgpt,
  title={Chatgpt for cybersecurity: practical applications, challenges, and future directions},
  author={Al-Hawawreh, Muna and Aljuhani, Ahamed and Jararweh, Yaser},
  journal={Cluster Computing},
  volume={26},
  number={6},
  pages={3421--3436},
  year={2023},
  publisher={Springer}
}

@misc{de2024devaic,
  title={DeVAIC: A tool for security assessment of AI-generated code},
  author={De Luca, Roberta and Cotroneo, Domenico and Liguori, Pietro},
  year={2024}
}

\end{document}